\documentclass[aps,prl,twocolumn,superscriptaddress,showpacs,english]{revtex4-1}

\usepackage[T1]{fontenc}
\usepackage[latin9]{inputenc}
\usepackage{babel}
\usepackage{amsmath}
\usepackage{amssymb}
\usepackage{wasysym}
\usepackage{graphicx}
\usepackage{xcolor}

\definecolor{mydarkgreen}{rgb}{0.0,0.5,0.0}
\definecolor{friebrick}{rgb}{0.698,0.1333,0.1333}

\usepackage[linktocpage=true,
  colorlinks=true, 
  pdfborder={0 0 0},
  linkcolor=blue,
  citecolor=friebrick,
  filecolor=yellow,
  urlcolor=mydarkgreen,
  bookmarks,
  pdfauthor={},
]{hyperref}

\newcommand{\mpi}{Max-Planck Institut f\"ur Microstrukture Physics, Weinberg 2, 06120 Halle, Germany}
\newcommand{\mus}{$\mu^*$}
\newcommand{\tc}{T$_{\textmd C}$}
\newcommand{\ef}{E$_{\rm F}$}
\newcommand{\omlog}{$\omega_{\textmd log}$}
\newcommand{\sh}{H$_3$S}
\newcommand{\sd}{D$_3$S}
\newcommand{\seh}{H$_3$Se}
\newcommand{\tcscdft}{T$_{\rm c}^{\rm SCDFT}$}
\newcommand{\tcscdftph}{T$_{\rm c}^{\rm SCDFT,ph}$}
\newcommand{\tcADph}{T$_{\rm c}^{\rm AD,\mu^*=0}$}
\newcommand{\tcAD}{T$_{\rm c}^{\rm AD,\mu^*=0.1}$}
\newcommand{\gapMB}{$\Delta$(T=0)}
\newcommand{\cubic}{$Im$\={3}$m$ }

\begin{document}

\title{High temperature superconductivity in sulfur and selenium hydrides at high pressure}

\author{Jos\'e A. Flores-Livas}
\author{Antonio Sanna}
\author{E.~K.~U. Gross}

\affiliation{\mpi}
\date{\today}

\begin{abstract}
Due to its low atomic mass hydrogen is the most promising element to search for high-temperature phononic superconductors. 
However, metallic phases of hydrogen are only expected at extreme pressures (400\,GPa or higher). 
The measurement of a record superconducting critical temperature of 190\,K in a hydrogen-sulfur 
compound at 200\,GPa of pressure~\cite{Eremets_arxiv2014}, shows that metallization of 
hydrogen can be reached at significantly lower pressure by inserting it in the matrix of other elements.  
In this work we re-investigate the phase diagram and the superconducting properties of the H-S system by means of minima hopping method for structure prediction 
and Density Functional theory for superconductors. 
We also show that Se-H has a similar phase diagram as its sulfur counterpart as well as high superconducting critical temperature. 
We predict \seh\ to exceed 120\,K superconductivity at 100\,GPa.  
We show that both \seh\ and \sh, due to the critical temperature and peculiar electronic structure, present rather unusual superconducting properties.
\end{abstract}

\pacs{~}

\maketitle


Under high pressure conditions, insulating and semiconducting materials tend to become metallic,  
because, with increasing electronic density, the kinetic energy grows faster than the potential energy.
As metallicity is a necessary condition for superconductivity, superconductivity becomes more likely under pressure~\cite{Shimizu_EHP2005,superconductingelements_2005}. 
Wigner and Huntington~\cite{Wigner_JCP1935}, already in 1935 suggested the possibility of a 
metallic modification of hydrogen under very high pressures.  Ashcroft and Richardson predicted~\cite{Ashcroft_PRL1968,RichardsonAshcroft_PRL97} 
hydrogen to become metallic under pressure and also the possibility to be a high temperature superconductor. 
The high critical temperature (\tc) of hydrogen~\cite{Cudazzo_PRL2008,Zhang_H_SSC2007,Cudazzo_1_PRB2010} is a consequence of 
its low atomic mass leading to high energy vibrational modes and in turn to a large phase space available 
for electron-phonon scattering to induce superconductivity~\cite{BCS_1957}. 
However, the estimated pressure of metallization~\cite{pickard_structure_2007,Cudazzo_2_PRB2010} is beyond the 
current experimental capabilities and it has been a challenge to confirm this hypothesis~\cite{LeToullec2002,Eremets_NatMat2011,Hemley_PRL2012,HRussell_hydrogenJACS2014}.

It was only recently that hydrogen-rich compounds started to be explored as a 
way to decrease the tremendous pressure of metallization in pure hydrogen~\cite{Ashcroft_PRL2004}, essentially performing a chemical pre-compression.
The first system explored experimentally was silane (SiH$_4$)~\cite{Eremets_Science2008}. Soon after, many other pre-compressed hydrogen rich materials have been explored experimentally~\cite{chen_pressure-induced_2008,degtyareva_formation_2009,Hanfland_PRL2011,Strobel_PRL2011}
and theoretically~\cite{tse_novel_2007,Chen_PNAS2008,Kim_PNAS2008,Wang_PNAS2009,Yao_PNAS2010,gao_high-pressure_2010,Kim_PNAS2010,Li_PNAS2010,Zhou_PRB2012,LiYanmingMa_JCP2014,Yanming_JCP2014,Hooper_JPC-2014,Duan_SciRep2014}. 
The importance of a systematic search for a crystalline ground state has been put in evidence for disilane (SiH$_6$), 
where structures enthalpically higher lead to transition temperatures of the order of 130\,K. 
Interesting structures have been proved not to be the global minimum and for the correct ground state was found a rather 
moderate electron-phonon coupling and \tc\ of 25\,K~\cite{Disilane_JAFL}. In agreement with experimental evidence. 

Recently it was reported that sulfur hydride (SH$_2$), when pressurized, becomes metallic and superconducting.
For pressures above 180\,GPa an extremely high transition temperature of  about 190\,K was measured~\cite{Eremets_arxiv2014}. 
This \tc\ is higher than in other superconductors known so far, including cuprates and pnictides.
The experimental evidence is supported theoretically~\cite{Duan_SciRep2014,LiYanmingMa_JCP2014,Bernstein_arxiv_2014}, 
and crystal prediction methods suggest that the system becomes superconducting with a SH$_3$ stoichiometry.
In this work  we re-investigate extensively the S-H phases with state of the art \textit{ab-initio} material search minima hopping methods~\cite{Goedecker_2004,Goedecker_2005,Amsler_2010} (MHM) and compute the superconducting properties with the completely parameter-free Density Functional Theory for Superconductors (SCDFT). 
We also extend the analysis to the Se-H system, predicting a fairly similar phase diagram and comparable superconducting properties. 

\begin{figure*}[ht]
\begin{center}
\begin{minipage}{0.49\textwidth}
\includegraphics[width=0.99\columnwidth,angle=0]{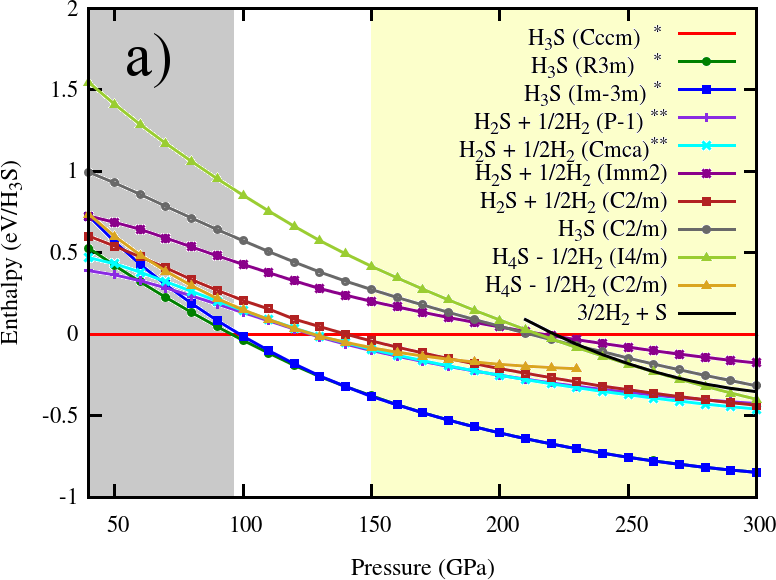}
\end{minipage}
\begin{minipage}{0.49\textwidth}
\includegraphics[width=0.99\columnwidth,angle=0]{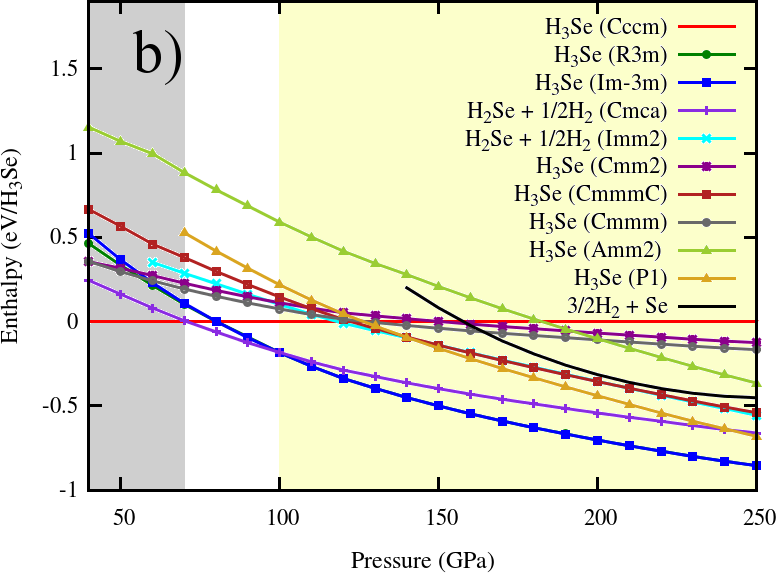}
\end{minipage}
\caption{(Color online) Calculated enthalpies for \sh\ (a) and \seh\ (b) structures and their decompositions.
 Values are given with respect to the $Cccm$ structure, stable at low pressure. In \sh\, three structures were already reported in 
 Ref.~\onlinecite{Duan_SciRep2014} (*) and two in Ref.~\onlinecite{LiYanmingMa_JCP2014} (**). The other structures of \sh\ as well as all the structures of \seh\ are predictions of this work by means of the Minima Hopping Method.}
 \label{fig:enthalpy}
\end{center}
\end{figure*}

\section{Methods}
Electronic and phononic structure calculations are performed within density-functional theory as implemented in the two plane-wave based codes {\sc abinit}~\cite{gonze_abinit_2009}, and {\sc espresso}~\cite{QE-2009}  within the local density approximation LDA exchange correlation functional. The core states were accounted for by norm-conserving Troullier-Martins pseudopotentials~\cite{FHI_Fuchs}. 
The pseudopotential accuracy has been checked against all-electron (LAPW+lo) method as implemented in the {\sc elk} code (http://elk.sourceforge.net/). 
In order to predict the ground state structure of sulfur/selenium hydride compounds we use the minima hopping method~\cite{Goedecker_2004,Goedecker_2005,Amsler_2010} for the prediction of low-enthalpy structures.  
This method has been successfully used for global geometry optimization in a large variety of applications~\cite{MA_JAFL,LiAlH_Maxmotif,BJAFL_PRB2012}. 
Given only the chemical composition of a system, the MHM aims at finding the global minimum on 
the enthalpy surface while gradually exploring low-lying structures.
Moves on the enthalpy surface are performed by using variable cell 
shape molecular dynamics with initial velocities approximately 
chosen along soft mode directions. We have used 1,2,3 formula units of \sh\ and \seh\ at selected 
pressures of 100, 150, 200, 250 and 300\,GPa. The relaxations to local minima are performed by the fast inertia relaxation engine~\cite{FIRE_2006} and both atomic and cell degrees of freedom are taking into account. 
Final structural relaxations and enthalpy calculations were performed with the {\sc vasp} code~\cite{VASP_Kresse}.  
The plane-wave cutoff energy was set to 800\,eV, and Monkhorst-Pack $k$-point meshes with grid spacing denser 
 than $2\pi\times 0.01$\,\AA$^{-1}$, resulting in total energy convergence to better than 1\,meV/atom. 
Superconducting properties have been computed within density-functional theory for superconductors (SCDFT)~\cite{OGK_SCDFT_PRL1988,Lueders_SCDFT_PRB2005,Marques_SCDFT_PRB2005}. This theory of superconductivity is completely \textit{ab-initio}, fully parameter-free and proved to be rather accurate and successful in describing phononic superconductors\cite{Floris_Pb_PRB2007,Gonnelli_CaC6_PRL2008,Sanna_CaC6_RapCom2007,Profeta_LiKAl_PRL2006}. It allows to compute all superconducting properties including the critical temperature and the excitation spectrum of the system~\footnote{The phononic functional we use is an improved version with respect to Ref.~\cite{Lueders_SCDFT_PRB2005,Marques_SCDFT_PRB2005} and is discussed in Ref.~\cite{Sanna_Migdal}. In this work Coulomb interactions are included within static RPA~\cite{Sanna_CaC6_RapCom2007}, therefore excluding magnetic source of coupling\cite{Frank_SF_PRB2014}}.

\section{Crystal structure prediction and enthalpies}

Experimentally little is known on the high pressure stability and composition of the S-H system and, to the best of our knowledge, nothing is known about the Se-H. Therefore we investigate their low temperature phase diagram by means of the MHM for the prediction of low-enthalpy structures.
Computed enthalpies as a function of pressure are reported in Fig.~\ref{fig:enthalpy}.  
We consider the H$_3$S stoichiometry as well as its elemental decomposition (sulfur + hydrogen), its decomposition into H$_2$S + hydrogen and H$_4$S - hydrogen~\footnote{
The decomposition enthalpies have been computed from the predicted structures of hydrogen $P6_3m$ and $C2/c$~\cite{pickard_structure_2007} 
and for sulfur and selenium on the $R3m$ and \cubic reported to occur at high pressure.~\cite{Akahama_S_PRB1993,Akahama_Se_PRB1993,Akahama_Se_met_PRB1997,Shimizu_EHP2005}}. 
At low pressure we find the $Cccm$ structure up to 95\,GPa and the $R3m$ ($\beta$-Po-type) rhombohedral structure between 95 and 150\,GPa. Above 150\,GPa, we confirm\cite{LiYanmingMa_JCP2014,Bernstein_arxiv_2014} the cubic \cubic (bcc) as the most stable lattice. 

In a similar way we have studied the Se-H phase diagram. Chemically, selenium is known to have very similar 
physical properties to sulfur and this system is not an exception. 
The enthalpies of the phases found in our MHM runs are shown in Fig.~\ref{fig:enthalpy}b. Once again we use the $Cccm$ structure as reference since, as in the case of sulfur, it is the most stable at low pressures and up to 70\,GPa. Between 70\,GPa to 100\,GPa, we find that the H$_2$Se + hydrogen decomposition is more stable than the \seh\ stoichiometry. \seh\ returns to be the most stable composition above 100\,GPa and at least up to 250\,GPa.  

Therefore from our analysis both systems in the range 50\,GPa to 250\,GPa show, with increasing pressure, two phase transitions. The S-H system, always stable in the \sh\ stoichiometry, has a first order phase transition from $Cccm$ to $R3m$ at $\sim$100\,GPa,  then the $R3m$ rhombohedral distortion decreases continuously up to 150\,GPa where it transforms into the \cubic cubic structure. 
The Se-H system at low pressure is also stable in the \seh\ stoichiometry but becomes unstable to a phase separation into H$_2$S + hydrogen in the range from 70\,GPa to 100\,GPa. Above 100\,GPa another discontinuous phase transition occurs, directly into the  \cubic cubic structure. Note that 100\,GPa is also the pressure below which the \cubic structure would distort into the $\beta$-Po $R3m$, therefore depending on experimental conditions this rhombohedral phase may occur as a metastable one. 
The sequence of transformation is highlighted in Fig.~\ref{fig:enthalpy} by means of shaded areas.

\section{electronic and phononic properties of \seh\ and \sh\ at high pressure}\label{sec:bandsandcoupling}

\begin{figure}[t]
\begin{center}
\includegraphics[width=0.97\columnwidth,angle=0]{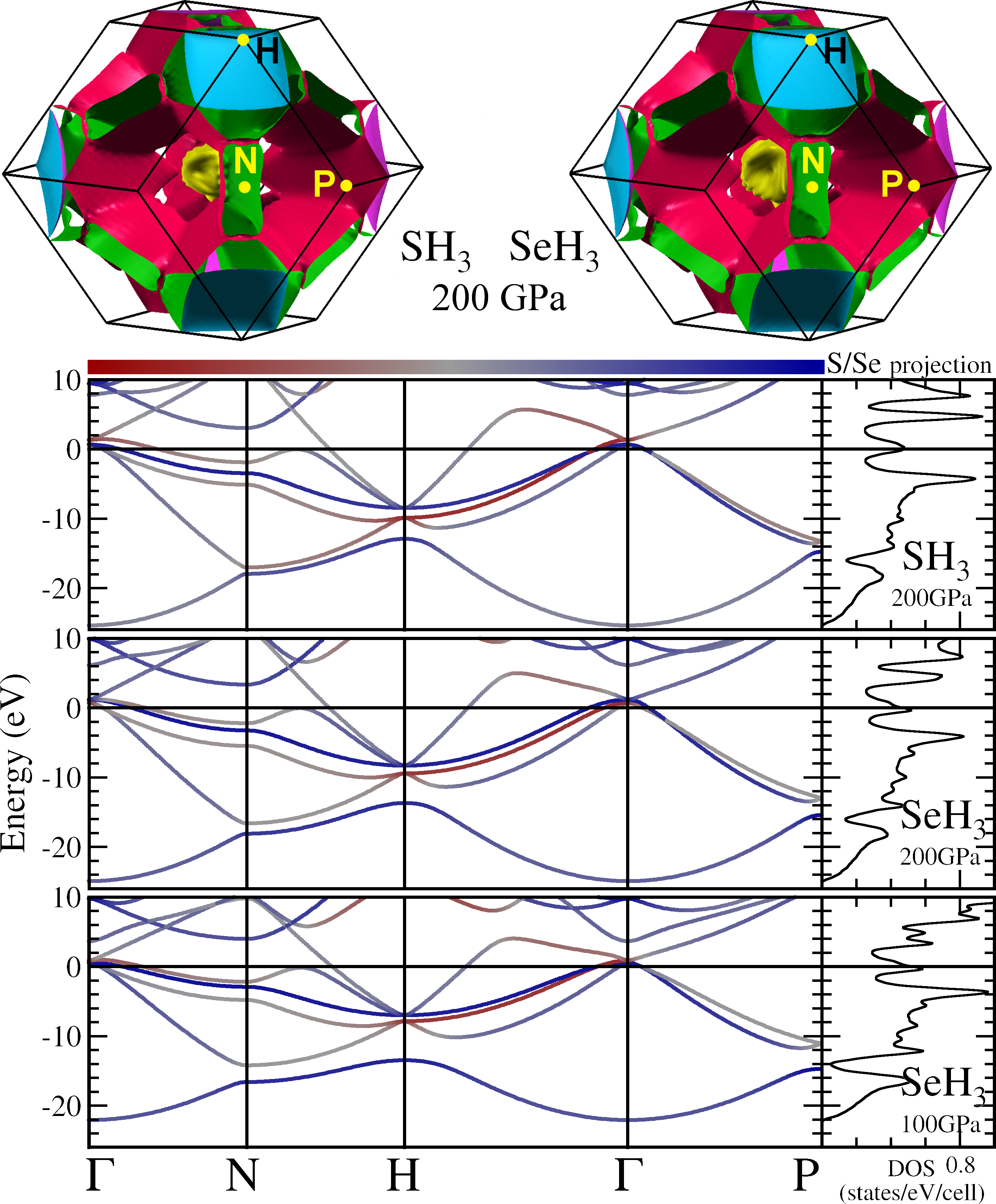}
\caption{(Color online) Fermi Surfaces  (top) and electronic band structures (bottom) of \sh\ and \seh\ at high pressure in the \cubic phase.
The color-scale in the band lines indicates the projection of the KS states on the atomic orbitals of the  sulfur/selenium atom
normalized by the maximum total atomic projection of these valence states that is of about 70\%.} 
 \label{fig:bands_and_fs}
\end{center}
\end{figure}

We focus now on the properties of \sh\ and \seh\ in the pressure range of stability of the \cubic structure.
The two materials present very similar properties. 
At 200\,GPa electronic band dispersions and Fermi surfaces are barely distinguishable, as seen in Fig.~\ref{fig:bands_and_fs}.
And in the range of pressure between 100 to 200\,GPa  there are no significant changes in the electronic properties 
apart from the overall bandwidth that increases with pressure.  
An important aspect of the electronic structure is the presence of several Fermi surface sheets, 
with no marked nesting features and with Fermi states both at low and high momentum vector.  
At small momentum (close to the $\Gamma$-point, center of the Brillouin zone in Fig.~\ref{fig:bands_and_fs}) there are three small Fermi surfaces 
(only the green larger one can be seen in the figure, smaller ones being inside it). However, these provide a rather small contribution 
to the total density of states (DOS) at the Fermi level which mostly comes from the two larger Fermi surface sheets. 
These are of \textit{hybrid} character, meaning that their Kohn-Sham (KS) states overlap both with H and S/Se states 
(the overlap is expressed in the figure by the color-scale of the band lines), 
 suggesting that they will be coupled with both hydrogen and S/Se lattice vibrations (more details on this point will be given in the next section).  
Overall the DOS shows a square root behavior of the 3D electron gas, the main deviation from this 
occurs close to the Fermi energy where a peak with an energy width of about 2\,eV is present. 
This structure will play a relevant role in the superconducting properties.

\begin{figure}[t]
\includegraphics[width=1.0\columnwidth,angle=0]{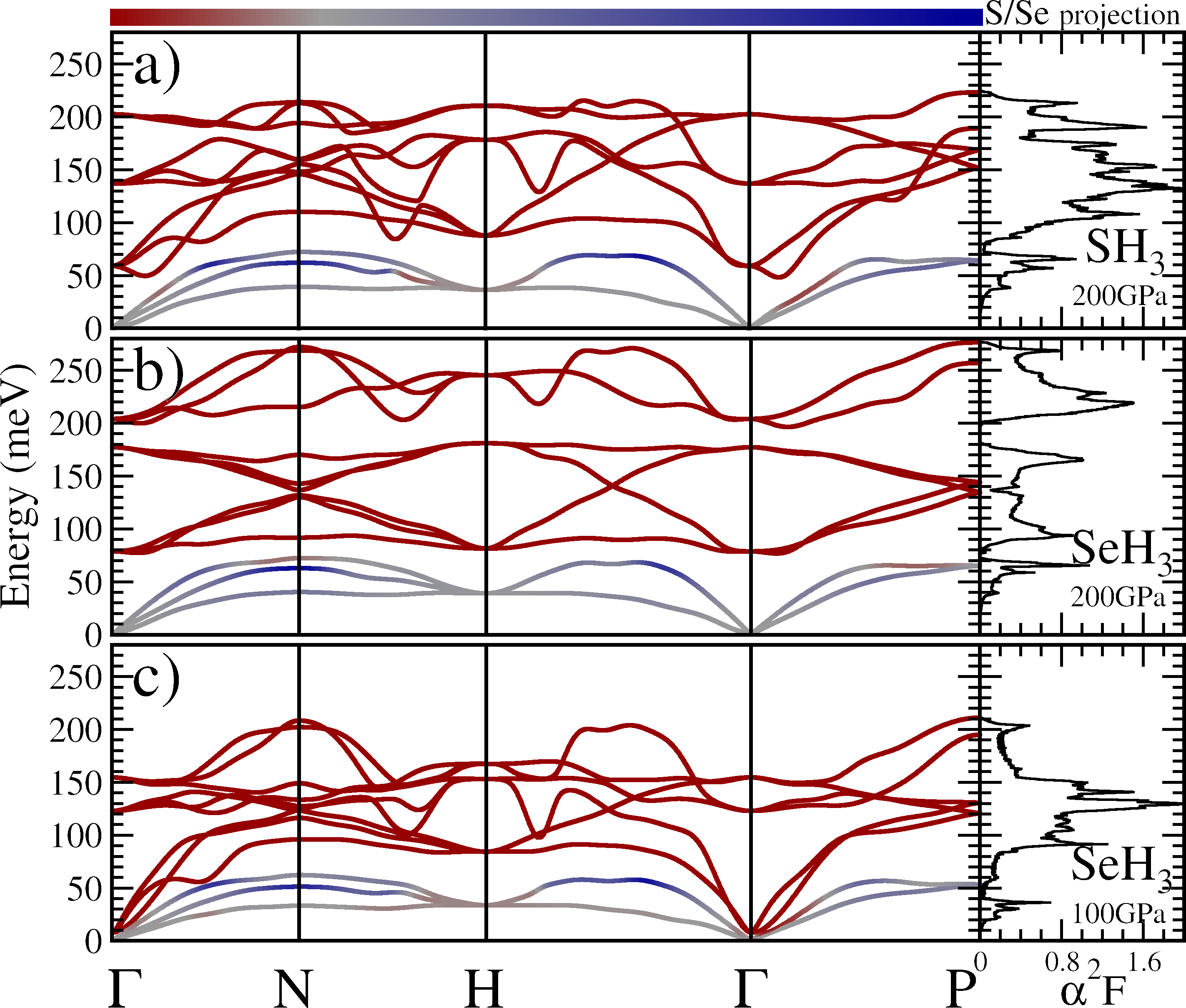}
  \caption{(Color online) Phonon dispersion and $\alpha^2F$ functions~\cite{Carbotte_RMP1990,AllenMitrovic1983} of \sh\ and \seh\ at high pressure in the cubic \cubic structure. The color coding gives the projection of the mode displacement on the  S/Se atom. 
  Displacements are visible dominated by H due to its lighter mass.}
 \label{fig:phonons}
\end{figure}

Unlike the electronic structure, phonons are strongly pressure and material dependent. Clearly a key role is played by the occurrence of the II order $R3m$ to \cubic phase transition. 
Far away from it (i.e. at very high pressure)  we have three sets of well separated phonon modes: acoustic (below 60\,meV), optical modes that are transverse with respect to the S/Se-H bond (between 100 to 200\,meV) and, above 200\,meV, stretching modes of the Se/S-H bond. 
These are clearly seen in Fig.~\ref{fig:phonons}b for \seh\ at 200\,GPa. 
As pressure reduces, the bond structure of the system tends to destabilize because, from a four-fold coordination in the \cubic  structure, it goes to a three-fold coordination in the $R3m$ one. This means that one of the high-energy stretching mode slowly softens at $\Gamma$. 
This softening can be clearly seen in \sh\ at 200\,GPa (Fig.~\ref{fig:phonons}a) where a H-S  stretching mode went down to about 60\,meV. 
Eventually, as pressure lowers this softens to zero energy, marking the occurrence of the phase transition, at about 150\,GPa in \sh\ and slightly below 100\,GPa in \seh. In fact, at 100\,GPa this mode has, in \seh, almost zero energy (see Fig.~\ref{fig:phonons}c).

In spite of these important changes in the phononic energy dispersion, the overall coupling strength\cite{Carbotte_RMP1990,AllenMitrovic1983} $\lambda$  does not show large variations over the pressure range, as we can see from Tab.~\ref{tab:tc}. Naturally the coupling increases near the phase transition due the optical mode softening, however, as this is restricted to a relatively small region near the $\Gamma$ point, the effect is not dramatic. On the other hand there is definitively a difference in the coupling strength of the Se ($\lambda \sim 1.5$)  with respect to the S system ($\lambda \sim 2.5$), indicating that selenium, due to its larger ionic size, provides a better electronic screening of the hydrogen vibrations.

\section{Superconducting properties}\label{sec:sc}

We have computed, by means of SCDFT, critical temperatures of \sh\ and \seh\ in the pressure range of stability of the \cubic structure, these are collected in Tab.~\ref{tab:tc}.
The predicted \tc\ for the \sh\ system is 180\,K at 200\,GPa and 195\,K at 180\,GPa, in agreement with the measured value of 185\,K at 177\,GPa.
On the other hand our prediction for the deuterium substituted system \sd\ is 141\,K, at 200\,GPa.  That is much larger than the measured\cite{Eremets_arxiv2014} \tc\ of 90\,K. This huge experimental isotope effect is therefore not consistent with our calculations. However the good agreement obtained with the \tc\ of the \sh\ system seem to exclude an explanation in terms of anharmonic effects in the hydrogen vibrations, as suggested in Ref.~\onlinecite{Pickett_Arxiv2015}.
Nevertheless, the theoretical isotope coefficients $\alpha^{S}=0.05$ and $\alpha^H=0.4$ (defined as $\alpha^A=-\frac{M^A}{Tc}\frac{\partial T_C}{\partial M^A}$, and computed at 200\,GPa with a three point numerical differentiation) clearly indicate and confirm~\cite{Pickett_Arxiv2015} the dominant contribution of hydrogen phonon modes to the superconducting pairing.

Our prediction for \seh\ at 200\,GPa is of 131\,K, this reduction of \tc\ is clearly not an isotopic effect of the 
substitution S to Se. As mentioned in the previous section, it is caused, instead, by a different coupling strength of the hydrogen modes in the Se environment.  
In spite of the much lower coupling strength $\lambda$ the reduction of \tc\ is not very large with respect to the sulfur system, as expected from the fact that the critical temperature at high coupling increases with the square root of $\lambda$ (while is exponential at low coupling)~\cite{Carbotte_RMP1990,AllenMitrovic1983}. 

To compute the critical temperatures we have used SCDFT, this choice allow us to deal 
with the unusual superconducting properties of these systems from first principles  without relying on conventional assumptions coming 
from low pressure experience. There are two aspects of these systems that are uncommon, that make the use of conventional~\footnote{Conventional implementation of the Eliashberg equations due to their computational cost, usually assume a $k$-independent pairing and a flat density of states. Anisotropic implementations \cite{Margine_anisoEliashberg_PRB2013} are not intrinsically affected by this limit.} 
Eliashberg methods difficult to apply to these systems. First, the strong variation of the electronic 
density of states at the Fermi level, that is pinned to a rather sharp peak in the DOS, 
second the extremely large el-ph coupling and phonon frequencies that lead to a very broad region around 
\ef\ where the interaction is dominated by phonons over Coulomb repulsion.

\setlength{\tabcolsep}{8pt}
\begin{table*}[ht]
\begin{tabular}{r r||c|c|c|c|c|c|c}
      &               &$\lambda$  &\omlog  & \tcscdft     &     \gapMB &  \tcscdftph &  \tcAD   & \tcADph  \\\hline\hline
\sh\  &{\small 200\,GPa}& 2.41    &109\,meV&{\bf 180\,K}  & 43.8\,meV  &   284\,K    &  255\,K  &  338\,K  \\
\sd\  &{\small 200\,GPa}& 2.41    & 82\,meV&{\bf 141\,K}  & 32.9\,meV  &   216\,K    &  188\,K  &  247\,K  \\
\sh\  &{\small 180\,GPa}& 2.57    &101\,meV&{\bf 195\,K}  & 44.8\,meV  &   297\,K    &  250\,K  &  331\,K  \\\hline
\seh\ &{\small 200\,GPa}& 1.45    &120\,meV&{\bf 131\,K}  & 28.4\,meV  &   234\,K    &  174\,K  &  246\,K  \\
\seh\ &{\small 150\,GPa}& 1.38    &107\,meV&{\bf 110\,K}  & 23.4\,meV  &   195\,K    &  145\,K  &  209\,K  \\
\seh\ &{\small 100\,GPa}& 1.76    & 87\,meV&{\bf 123\,K}  & 27.0\,meV  &   198\,K    &  156\,K  &  214\,K  \\
\end{tabular}
\caption{Calculated critical Temperatures and gaps.
$\lambda$ is the electron phonon coupling parameter~\cite{Carbotte_RMP1990,AllenMitrovic1983};
\omlog\ is the logarithmic average of the $\alpha^2F$ function~\cite{Carbotte_RMP1990,AllenMitrovic1983};
\tcscdft\ is the critical temperature from SCDFT including RPA screened Coulomb repulsion; 
\tcscdftph\ is the phonon only SCDFT critical temperature; 
\tcADph\ is the critical temperature from the Allen-Dynes modified McMillan formula~\cite{AllenDynes_PRB1975,McMillanTc,AllenMitrovic1983}, 
at $\mu^*=0$ (i.e. with no Coulomb pairing);
\tcAD\ is the same AD formula but with the conventional value of $\mu^*=0.1$ ;
\gapMB\ is the superconducting gap computed~\cite{Sanna_Migdal} from the SCDFT calculations}\label{tab:tc}
\end{table*}

The effect of the energy dependence of the DOS can be appreciated by comparing Eliashberg results with SCDFT 
when neglecting the Coulomb interaction (see Tab.~\ref{tab:tc}). 
At 200\,GPa the two theories~\footnote{We are actually not reporting the Eliashberg result but that coming from the Allen-Dynes (AD) formula. 
The reason for this choice is that the two approaches agree perfectly (the difference being less than 1\,K) for the phonon case. 
But in addition when including \mus\ the AD formula depends only on it, while the Eliashberg equations also 
depend on the Coulomb frequency cut-off (that changes the meaning of the $*$ in \mus. If we want to use a 
conventional value of \mus\ between 0.1 and 0.15~\cite{AllenMitrovic1983} it is then better to use the 
parametrized AD version of the Eliashberg method} disagree by 54\,K, SCDFT giving 284\,K while Eliashberg gives 338\,K. 
This difference comes from the energy dependence of the DOS, while Eliashberg assumes a flat DOS, in the 
SCDFT we can easily check this assumption by assuming a flat DOS, and for this case the SCDFT calculation 
would lead to 334\,K, in agreement with the Eliashberg result. 

Physically the reduction of \tc, occurring when the real DOS is considered, arises from the fact that the 
phononic pairing extends in a rather large region around the Fermi level, over the DOS peak structure of these systems (see Fig.~\ref{fig:bands_and_fs}).
Beyond the range of the phononic pairing the coupling is dominated by the Coulomb interaction. As, in the static limit, this is repulsive, a superconducting system compensates it by a phase shift in the gap (i.e. in the quasiparticle orbitals), therefore making this repulsion contribute to the condensation (in unconventional superconductors exactly the same happens but directly at the Fermi level). This mechanism is called Coulomb renormalization~\cite{AllenMitrovic1983} since it \textit{renormalizes} the repulsive Coulomb scattering that occurs at low energy.
The phase shift occurs at $|\epsilon_{\bf k}|\gtrsim$\omlog\ but the scattering processes become less and less important as $|\epsilon_{\bf k}|$ increases (going down as 1/$\epsilon$). Therefore the most important energy region is where the DOS of the \sh\ and \seh\ systems shows a dip, implying that the phase space available for this process is small and its effect weak. 
Note that in order to reproduce the \tc\ coming from SCDFT within the Allen-Dynes (AD) formula one should assume a \mus\ of 0.16, that is actually much larger than the value of  $\mu$ itself ($\simeq$0.1). Making clear that the the Morel-Anderson theory\cite{MorelAnderson_1962} can not even be applied.
 
The superconducting pairing is distributed over many phonon modes and over the Brillouin zone in ${\bf q}$-space, 
despite the presence of several Fermi surface sheets and with different orbital character across the Fermi level, 
we obtained a isotropic (weakly ${\bf k}$-dependent) gap at the Fermi level 
and the effect of anisotropy~\cite{SMW_multibandBCS_PRL1959} on \tc\ is negligible ($<$ 1\,K).

\section{Conclusions}

We have presented a theoretical investigation on the crystal structure and superconductivity of \sh.  
An extensive structural search confirms the \sh\ stoichiometry as the most stable configuration at high pressure.   
By means of parameter-free SCDFT we have predicted a \tc\ of 180\,K at 200\,GPa, in excellent agreement with experimental results. 
This confirms \sh\ to be the material with the highest known superconducting critical temperature.
The mechanism of superconductivity is clearly the same that was predicted for metallic hydrogen\cite{Ashcroft_PRL1968,Cudazzo_PRL2008,Cudazzo_1_PRB2010}: the combinate effect of high characteristic frequency due to hydrogen light mass and strong coupling due to the lack of electronic core in hydrogen.
Still the working pressures of this superconductor is too high for any technological application~\cite{PaulMc_NatHPMat}. Nevertheless the discovery of metallic superconducting hydrogenic bands already at 150\,GPa gives hope that further theoretical and experimental research in this direction 
may lead to even lower hydrogen metallization pressures and higher temperature superconductivity.
Here we predict that \sh\ is stable in the cubic \cubic already at 100\,GPa with a very high \tc\ of 123\,K, a value which is comparable to the cuprate superconductors.

\begin{acknowledgments}
J.A.F.L. acknowledge fruitful discussion with Maximilian Amsler on crystal prediction and 
the financial support from EU's 7th framework Marie-Sk\l{}odowska-Curie scholarship program within 
the ``ExMaMa'' Project (329386) 
\end{acknowledgments} 

\bibliographystyle{apsrev4-1}
\bibliography{paper}

\begin{thebibliography}{73}%
\makeatletter
\providecommand \@ifxundefined [1]{%
 \@ifx{#1\undefined}
}%
\providecommand \@ifnum [1]{%
 \ifnum #1\expandafter \@firstoftwo
 \else \expandafter \@secondoftwo
 \fi
}%
\providecommand \@ifx [1]{%
 \ifx #1\expandafter \@firstoftwo
 \else \expandafter \@secondoftwo
 \fi
}%
\providecommand \natexlab [1]{#1}%
\providecommand \enquote  [1]{``#1''}%
\providecommand \bibnamefont  [1]{#1}%
\providecommand \bibfnamefont [1]{#1}%
\providecommand \citenamefont [1]{#1}%
\providecommand \href@noop [0]{\@secondoftwo}%
\providecommand \href [0]{\begingroup \@sanitize@url \@href}%
\providecommand \@href[1]{\@@startlink{#1}\@@href}%
\providecommand \@@href[1]{\endgroup#1\@@endlink}%
\providecommand \@sanitize@url [0]{\catcode `\\12\catcode `\$12\catcode
  `\&12\catcode `\#12\catcode `\^12\catcode `\_12\catcode `\%12\relax}%
\providecommand \@@startlink[1]{}%
\providecommand \@@endlink[0]{}%
\providecommand \url  [0]{\begingroup\@sanitize@url \@url }%
\providecommand \@url [1]{\endgroup\@href {#1}{\urlprefix }}%
\providecommand \urlprefix  [0]{URL }%
\providecommand \Eprint [0]{\href }%
\providecommand \doibase [0]{http://dx.doi.org/}%
\providecommand \selectlanguage [0]{\@gobble}%
\providecommand \bibinfo  [0]{\@secondoftwo}%
\providecommand \bibfield  [0]{\@secondoftwo}%
\providecommand \translation [1]{[#1]}%
\providecommand \BibitemOpen [0]{}%
\providecommand \bibitemStop [0]{}%
\providecommand \bibitemNoStop [0]{.\EOS\space}%
\providecommand \EOS [0]{\spacefactor3000\relax}%
\providecommand \BibitemShut  [1]{\csname bibitem#1\endcsname}%
\let\auto@bib@innerbib\@empty
\bibitem [{\citenamefont {Drozdov}\ \emph {et~al.}(2014)\citenamefont
  {Drozdov}, \citenamefont {Eremets},\ and\ \citenamefont
  {Troyan}}]{Eremets_arxiv2014}%
  \BibitemOpen
  \bibfield  {author} {\bibinfo {author} {\bibfnamefont {A.}~\bibnamefont
  {Drozdov}}, \bibinfo {author} {\bibfnamefont {M.~I.}\ \bibnamefont
  {Eremets}}, \ and\ \bibinfo {author} {\bibfnamefont {I.~A.}\ \bibnamefont
  {Troyan}},\ }\href {arXiv:1412.0460 [cond-mat.supr-con]} {\bibfield
  {journal} {\bibinfo  {journal} {arXiv:1412.0460 [cond-mat.supr-con]}\ }
  (\bibinfo {year} {2014})}\BibitemShut {NoStop}%
\bibitem [{\citenamefont {Shimizu}\ \emph {et~al.}(2005)\citenamefont
  {Shimizu}, \citenamefont {Amaya},\ and\ \citenamefont
  {Suzuki}}]{Shimizu_EHP2005}%
  \BibitemOpen
  \bibfield  {author} {\bibinfo {author} {\bibfnamefont {K.}~\bibnamefont
  {Shimizu}}, \bibinfo {author} {\bibfnamefont {K.}~\bibnamefont {Amaya}}, \
  and\ \bibinfo {author} {\bibfnamefont {N.}~\bibnamefont {Suzuki}},\ }\href
  {\doibase 10.1143/JPSJ.74.1345} {\bibfield  {journal} {\bibinfo  {journal}
  {Journal of the Physical Society of Japan}\ }\textbf {\bibinfo {volume}
  {74}},\ \bibinfo {pages} {1345} (\bibinfo {year} {2005})},\ \Eprint
  {http://arxiv.org/abs/http://dx.doi.org/10.1143/JPSJ.74.1345}
  {http://dx.doi.org/10.1143/JPSJ.74.1345} \BibitemShut {NoStop}%
\bibitem [{\citenamefont {Buzea}\ and\ \citenamefont
  {Robbie}(2005)}]{superconductingelements_2005}%
  \BibitemOpen
  \bibfield  {author} {\bibinfo {author} {\bibfnamefont {C.}~\bibnamefont
  {Buzea}}\ and\ \bibinfo {author} {\bibfnamefont {K.}~\bibnamefont {Robbie}},\
  }\href {http://stacks.iop.org/0953-2048/18/i=1/a=R01} {\bibfield  {journal}
  {\bibinfo  {journal} {Superconductor Science and Technology}\ }\textbf
  {\bibinfo {volume} {18}},\ \bibinfo {pages} {R1} (\bibinfo {year}
  {2005})}\BibitemShut {NoStop}%
\bibitem [{\citenamefont {Wigner}\ and\ \citenamefont
  {Huntington}(1935)}]{Wigner_JCP1935}%
  \BibitemOpen
  \bibfield  {author} {\bibinfo {author} {\bibfnamefont {E.}~\bibnamefont
  {Wigner}}\ and\ \bibinfo {author} {\bibfnamefont {H.~B.}\ \bibnamefont
  {Huntington}},\ }\href {\doibase http://dx.doi.org/10.1063/1.1749590}
  {\bibfield  {journal} {\bibinfo  {journal} {The Journal of Chemical Physics}\
  }\textbf {\bibinfo {volume} {3}},\ \bibinfo {pages} {764} (\bibinfo {year}
  {1935})}\BibitemShut {NoStop}%
\bibitem [{\citenamefont {Ashcroft}(1968)}]{Ashcroft_PRL1968}%
  \BibitemOpen
  \bibfield  {author} {\bibinfo {author} {\bibfnamefont {N.}~\bibnamefont
  {Ashcroft}},\ }\href {\doibase 10.1103/PhysRevLett.21.1748} {\bibfield
  {journal} {\bibinfo  {journal} {Phys. Rev. Lett.}\ }\textbf {\bibinfo
  {volume} {21}},\ \bibinfo {pages} {1748} (\bibinfo {year}
  {1968})}\BibitemShut {NoStop}%
\bibitem [{\citenamefont {Richardson}\ and\ \citenamefont
  {Ashcroft}(1997)}]{RichardsonAshcroft_PRL97}%
  \BibitemOpen
  \bibfield  {author} {\bibinfo {author} {\bibfnamefont {C.~F.}\ \bibnamefont
  {Richardson}}\ and\ \bibinfo {author} {\bibfnamefont {N.~W.}\ \bibnamefont
  {Ashcroft}},\ }\href {\doibase 10.1103/PhysRevLett.78.118} {\bibfield
  {journal} {\bibinfo  {journal} {Phys. Rev. Lett.}\ }\textbf {\bibinfo
  {volume} {78}},\ \bibinfo {pages} {118} (\bibinfo {year} {1997})}\BibitemShut
  {NoStop}%
\bibitem [{\citenamefont {Cudazzo}\ \emph {et~al.}(2008)\citenamefont
  {Cudazzo}, \citenamefont {Profeta}, \citenamefont {Sanna}, \citenamefont
  {Floris}, \citenamefont {Continenza}, \citenamefont {Massidda},\ and\
  \citenamefont {Gross}}]{Cudazzo_PRL2008}%
  \BibitemOpen
  \bibfield  {author} {\bibinfo {author} {\bibfnamefont {P.}~\bibnamefont
  {Cudazzo}}, \bibinfo {author} {\bibfnamefont {G.}~\bibnamefont {Profeta}},
  \bibinfo {author} {\bibfnamefont {A.}~\bibnamefont {Sanna}}, \bibinfo
  {author} {\bibfnamefont {A.}~\bibnamefont {Floris}}, \bibinfo {author}
  {\bibfnamefont {A.}~\bibnamefont {Continenza}}, \bibinfo {author}
  {\bibfnamefont {S.}~\bibnamefont {Massidda}}, \ and\ \bibinfo {author}
  {\bibfnamefont {E.}~\bibnamefont {Gross}},\ }\href {\doibase
  10.1103/PhysRevLett.100.257001} {\bibfield  {journal} {\bibinfo  {journal}
  {Phys. Rev. Lett.}\ }\textbf {\bibinfo {volume} {100}},\ \bibinfo {pages}
  {257001} (\bibinfo {year} {2008})}\BibitemShut {NoStop}%
\bibitem [{\citenamefont {Zhang}\ \emph {et~al.}(2007)\citenamefont {Zhang},
  \citenamefont {Niu}, \citenamefont {Li}, \citenamefont {Cui}, \citenamefont
  {Wang}, \citenamefont {Ma}, \citenamefont {He},\ and\ \citenamefont
  {Zou}}]{Zhang_H_SSC2007}%
  \BibitemOpen
  \bibfield  {author} {\bibinfo {author} {\bibfnamefont {L.}~\bibnamefont
  {Zhang}}, \bibinfo {author} {\bibfnamefont {Y.}~\bibnamefont {Niu}}, \bibinfo
  {author} {\bibfnamefont {Q.}~\bibnamefont {Li}}, \bibinfo {author}
  {\bibfnamefont {T.}~\bibnamefont {Cui}}, \bibinfo {author} {\bibfnamefont
  {Y.}~\bibnamefont {Wang}}, \bibinfo {author} {\bibfnamefont {Y.}~\bibnamefont
  {Ma}}, \bibinfo {author} {\bibfnamefont {Z.}~\bibnamefont {He}}, \ and\
  \bibinfo {author} {\bibfnamefont {G.}~\bibnamefont {Zou}},\ }\href {\doibase
  http://dx.doi.org/10.1016/j.ssc.2006.12.029} {\bibfield  {journal} {\bibinfo
  {journal} {Solid State Communications}\ }\textbf {\bibinfo {volume} {141}},\
  \bibinfo {pages} {610 } (\bibinfo {year} {2007})}\BibitemShut {NoStop}%
\bibitem [{\citenamefont {Cudazzo}\ \emph
  {et~al.}(2010{\natexlab{a}})\citenamefont {Cudazzo}, \citenamefont {Profeta},
  \citenamefont {Sanna}, \citenamefont {Floris}, \citenamefont {Continenza},
  \citenamefont {Massidda},\ and\ \citenamefont {Gross}}]{Cudazzo_1_PRB2010}%
  \BibitemOpen
  \bibfield  {author} {\bibinfo {author} {\bibfnamefont {P.}~\bibnamefont
  {Cudazzo}}, \bibinfo {author} {\bibfnamefont {G.}~\bibnamefont {Profeta}},
  \bibinfo {author} {\bibfnamefont {A.}~\bibnamefont {Sanna}}, \bibinfo
  {author} {\bibfnamefont {A.}~\bibnamefont {Floris}}, \bibinfo {author}
  {\bibfnamefont {A.}~\bibnamefont {Continenza}}, \bibinfo {author}
  {\bibfnamefont {S.}~\bibnamefont {Massidda}}, \ and\ \bibinfo {author}
  {\bibfnamefont {E.~K.~U.}\ \bibnamefont {Gross}},\ }\href {\doibase
  10.1103/PhysRevB.81.134505} {\bibfield  {journal} {\bibinfo  {journal} {Phys.
  Rev. B}\ }\textbf {\bibinfo {volume} {81}},\ \bibinfo {pages} {134505}
  (\bibinfo {year} {2010}{\natexlab{a}})}\BibitemShut {NoStop}%
\bibitem [{\citenamefont {Bardeen}\ \emph {et~al.}(1957)\citenamefont
  {Bardeen}, \citenamefont {Cooper},\ and\ \citenamefont
  {Schrieffer}}]{BCS_1957}%
  \BibitemOpen
  \bibfield  {author} {\bibinfo {author} {\bibfnamefont {J.}~\bibnamefont
  {Bardeen}}, \bibinfo {author} {\bibfnamefont {L.~N.}\ \bibnamefont {Cooper}},
  \ and\ \bibinfo {author} {\bibfnamefont {J.~R.}\ \bibnamefont {Schrieffer}},\
  }\href {\doibase 10.1103/PhysRev.108.1175} {\bibfield  {journal} {\bibinfo
  {journal} {Phys. Rev.}\ }\textbf {\bibinfo {volume} {108}},\ \bibinfo {pages}
  {1175} (\bibinfo {year} {1957})}\BibitemShut {NoStop}%
\bibitem [{\citenamefont {Pickard}\ and\ \citenamefont
  {Needs}(2007)}]{pickard_structure_2007}%
  \BibitemOpen
  \bibfield  {author} {\bibinfo {author} {\bibfnamefont {C.~J.}\ \bibnamefont
  {Pickard}}\ and\ \bibinfo {author} {\bibfnamefont {R.~J.}\ \bibnamefont
  {Needs}},\ }\href {\doibase 10.1038/nphys625} {\bibfield  {journal} {\bibinfo
   {journal} {Nat Phys}\ }\textbf {\bibinfo {volume} {3}},\ \bibinfo {pages}
  {473} (\bibinfo {year} {2007})}\BibitemShut {NoStop}%
\bibitem [{\citenamefont {Cudazzo}\ \emph
  {et~al.}(2010{\natexlab{b}})\citenamefont {Cudazzo}, \citenamefont {Profeta},
  \citenamefont {Sanna}, \citenamefont {Floris}, \citenamefont {Continenza},
  \citenamefont {Massidda},\ and\ \citenamefont {Gross}}]{Cudazzo_2_PRB2010}%
  \BibitemOpen
  \bibfield  {author} {\bibinfo {author} {\bibfnamefont {P.}~\bibnamefont
  {Cudazzo}}, \bibinfo {author} {\bibfnamefont {G.}~\bibnamefont {Profeta}},
  \bibinfo {author} {\bibfnamefont {A.}~\bibnamefont {Sanna}}, \bibinfo
  {author} {\bibfnamefont {A.}~\bibnamefont {Floris}}, \bibinfo {author}
  {\bibfnamefont {A.}~\bibnamefont {Continenza}}, \bibinfo {author}
  {\bibfnamefont {S.}~\bibnamefont {Massidda}}, \ and\ \bibinfo {author}
  {\bibfnamefont {E.}~\bibnamefont {Gross}},\ }\href {\doibase
  10.1103/PhysRevB.81.134506} {\bibfield  {journal} {\bibinfo  {journal} {Phys.
  Rev. B}\ }\textbf {\bibinfo {volume} {81}},\ \bibinfo {pages} {134506}
  (\bibinfo {year} {2010}{\natexlab{b}})}\BibitemShut {NoStop}%
\bibitem [{\citenamefont {Loubeyre}\ \emph {et~al.}(2002)\citenamefont
  {Loubeyre}, \citenamefont {Occelli},\ and\ \citenamefont
  {LeToullec}}]{LeToullec2002}%
  \BibitemOpen
  \bibfield  {author} {\bibinfo {author} {\bibfnamefont {P.}~\bibnamefont
  {Loubeyre}}, \bibinfo {author} {\bibfnamefont {F.}~\bibnamefont {Occelli}}, \
  and\ \bibinfo {author} {\bibfnamefont {R.}~\bibnamefont {LeToullec}},\ }\href
  {\doibase 10.1038/416613a} {\bibfield  {journal} {\bibinfo  {journal}
  {Nature}\ }\textbf {\bibinfo {volume} {416}},\ \bibinfo {pages} {13}
  (\bibinfo {year} {2002})}\BibitemShut {NoStop}%
\bibitem [{\citenamefont {Eremets}\ and\ \citenamefont
  {Troyan}(2011)}]{Eremets_NatMat2011}%
  \BibitemOpen
  \bibfield  {author} {\bibinfo {author} {\bibfnamefont {M.~I.}\ \bibnamefont
  {Eremets}}\ and\ \bibinfo {author} {\bibfnamefont {I.~A.}\ \bibnamefont
  {Troyan}},\ }\href {\doibase http://dx.doi.org/10.1038/nmat3175} {\bibfield
  {journal} {\bibinfo  {journal} {Nat. Mat.}\ }\textbf {\bibinfo {volume}
  {10}},\ \bibinfo {pages} {927} (\bibinfo {year} {2011})}\BibitemShut
  {NoStop}%
\bibitem [{\citenamefont {Zha}\ \emph {et~al.}(2012)\citenamefont {Zha},
  \citenamefont {Liu},\ and\ \citenamefont {Hemley}}]{Hemley_PRL2012}%
  \BibitemOpen
  \bibfield  {author} {\bibinfo {author} {\bibfnamefont {C.-S.}\ \bibnamefont
  {Zha}}, \bibinfo {author} {\bibfnamefont {Z.}~\bibnamefont {Liu}}, \ and\
  \bibinfo {author} {\bibfnamefont {R.}~\bibnamefont {Hemley}},\ }\href
  {\doibase 10.1103/PhysRevLett.108.146402} {\bibfield  {journal} {\bibinfo
  {journal} {Phys. Rev. Lett.}\ }\textbf {\bibinfo {volume} {108}},\ \bibinfo
  {pages} {146402} (\bibinfo {year} {2012})}\BibitemShut {NoStop}%
\bibitem [{\citenamefont {Naumov}\ and\ \citenamefont
  {Hemley}(2014)}]{HRussell_hydrogenJACS2014}%
  \BibitemOpen
  \bibfield  {author} {\bibinfo {author} {\bibfnamefont {I.~I.}\ \bibnamefont
  {Naumov}}\ and\ \bibinfo {author} {\bibfnamefont {R.~J.}\ \bibnamefont
  {Hemley}},\ }\href {\doibase 10.1021/ar5002654} {\bibfield  {journal}
  {\bibinfo  {journal} {Accounts of Chemical Research}\ }\textbf {\bibinfo
  {volume} {47}},\ \bibinfo {pages} {3551} (\bibinfo {year} {2014})},\ \bibinfo
  {note} {pMID: 25369180},\ \Eprint
  {http://arxiv.org/abs/http://dx.doi.org/10.1021/ar5002654}
  {http://dx.doi.org/10.1021/ar5002654} \BibitemShut {NoStop}%
\bibitem [{\citenamefont {Ashcroft}(2004)}]{Ashcroft_PRL2004}%
  \BibitemOpen
  \bibfield  {author} {\bibinfo {author} {\bibfnamefont {N.}~\bibnamefont
  {Ashcroft}},\ }\href {\doibase 10.1103/PhysRevLett.92.187002} {\bibfield
  {journal} {\bibinfo  {journal} {Phys. Rev. Lett.}\ }\textbf {\bibinfo
  {volume} {92}},\ \bibinfo {pages} {187002} (\bibinfo {year}
  {2004})}\BibitemShut {NoStop}%
\bibitem [{\citenamefont {Eremets}\ \emph {et~al.}(2008)\citenamefont
  {Eremets}, \citenamefont {Trojan}, \citenamefont {Medvedev}, \citenamefont
  {Tse},\ and\ \citenamefont {Yao}}]{Eremets_Science2008}%
  \BibitemOpen
  \bibfield  {author} {\bibinfo {author} {\bibfnamefont {M.~I.}\ \bibnamefont
  {Eremets}}, \bibinfo {author} {\bibfnamefont {I.~A.}\ \bibnamefont {Trojan}},
  \bibinfo {author} {\bibfnamefont {S.~A.}\ \bibnamefont {Medvedev}}, \bibinfo
  {author} {\bibfnamefont {J.~S.}\ \bibnamefont {Tse}}, \ and\ \bibinfo
  {author} {\bibfnamefont {Y.}~\bibnamefont {Yao}},\ }\href {\doibase
  10.1126/science.1153282} {\bibfield  {journal} {\bibinfo  {journal}
  {Science}\ }\textbf {\bibinfo {volume} {319}},\ \bibinfo {pages} {1506}
  (\bibinfo {year} {2008})}\BibitemShut {NoStop}%
\bibitem [{\citenamefont {Chen}\ \emph
  {et~al.}(2008{\natexlab{a}})\citenamefont {Chen}, \citenamefont {Struzhkin},
  \citenamefont {Song}, \citenamefont {Goncharov}, \citenamefont {Ahart},
  \citenamefont {Liu}, \citenamefont {Mao},\ and\ \citenamefont
  {Hemley}}]{chen_pressure-induced_2008}%
  \BibitemOpen
  \bibfield  {author} {\bibinfo {author} {\bibfnamefont {X.}~\bibnamefont
  {Chen}}, \bibinfo {author} {\bibfnamefont {V.~V.}\ \bibnamefont {Struzhkin}},
  \bibinfo {author} {\bibfnamefont {Y.}~\bibnamefont {Song}}, \bibinfo {author}
  {\bibfnamefont {A.~F.}\ \bibnamefont {Goncharov}}, \bibinfo {author}
  {\bibfnamefont {M.}~\bibnamefont {Ahart}}, \bibinfo {author} {\bibfnamefont
  {Z.}~\bibnamefont {Liu}}, \bibinfo {author} {\bibfnamefont {H.-k.}\
  \bibnamefont {Mao}}, \ and\ \bibinfo {author} {\bibfnamefont {R.~J.}\
  \bibnamefont {Hemley}},\ }\href {\doibase 10.1073/pnas.0710473105} {\bibfield
   {journal} {\bibinfo  {journal} {Proceedings of the National Academy of
  Sciences}\ }\textbf {\bibinfo {volume} {105}},\ \bibinfo {pages} {20}
  (\bibinfo {year} {2008}{\natexlab{a}})}\BibitemShut {NoStop}%
\bibitem [{\citenamefont {Degtyareva}\ \emph {et~al.}(2009)\citenamefont
  {Degtyareva}, \citenamefont {Proctor}, \citenamefont {Guillaume},
  \citenamefont {Gregoryanz},\ and\ \citenamefont
  {Hanfland}}]{degtyareva_formation_2009}%
  \BibitemOpen
  \bibfield  {author} {\bibinfo {author} {\bibfnamefont {O.}~\bibnamefont
  {Degtyareva}}, \bibinfo {author} {\bibfnamefont {J.~E.}\ \bibnamefont
  {Proctor}}, \bibinfo {author} {\bibfnamefont {C.~L.}\ \bibnamefont
  {Guillaume}}, \bibinfo {author} {\bibfnamefont {E.}~\bibnamefont
  {Gregoryanz}}, \ and\ \bibinfo {author} {\bibfnamefont {M.}~\bibnamefont
  {Hanfland}},\ }\href {\doibase 10.1016/j.ssc.2009.07.022} {\bibfield
  {journal} {\bibinfo  {journal} {Solid State Communications}\ }\textbf
  {\bibinfo {volume} {149}},\ \bibinfo {pages} {1583} (\bibinfo {year}
  {2009})}\BibitemShut {NoStop}%
\bibitem [{\citenamefont {Hanfland}\ \emph {et~al.}(2011)\citenamefont
  {Hanfland}, \citenamefont {Proctor}, \citenamefont {Guillaume}, \citenamefont
  {Degtyareva},\ and\ \citenamefont {Gregoryanz}}]{Hanfland_PRL2011}%
  \BibitemOpen
  \bibfield  {author} {\bibinfo {author} {\bibfnamefont {M.}~\bibnamefont
  {Hanfland}}, \bibinfo {author} {\bibfnamefont {J.~E.}\ \bibnamefont
  {Proctor}}, \bibinfo {author} {\bibfnamefont {C.~L.}\ \bibnamefont
  {Guillaume}}, \bibinfo {author} {\bibfnamefont {O.}~\bibnamefont
  {Degtyareva}}, \ and\ \bibinfo {author} {\bibfnamefont {E.}~\bibnamefont
  {Gregoryanz}},\ }\href {\doibase 10.1103/PhysRevLett.106.095503} {\bibfield
  {journal} {\bibinfo  {journal} {Phy. Rev. Lett.}\ }\textbf {\bibinfo {volume}
  {106}},\ \bibinfo {pages} {095503} (\bibinfo {year} {2011})}\BibitemShut
  {NoStop}%
\bibitem [{\citenamefont {Strobel}\ \emph {et~al.}(2011)\citenamefont
  {Strobel}, \citenamefont {Ganesh}, \citenamefont {Somayazulu}, \citenamefont
  {Kent},\ and\ \citenamefont {Hemley}}]{Strobel_PRL2011}%
  \BibitemOpen
  \bibfield  {author} {\bibinfo {author} {\bibfnamefont {T.}~\bibnamefont
  {Strobel}}, \bibinfo {author} {\bibfnamefont {P.}~\bibnamefont {Ganesh}},
  \bibinfo {author} {\bibfnamefont {M.}~\bibnamefont {Somayazulu}}, \bibinfo
  {author} {\bibfnamefont {P.}~\bibnamefont {Kent}}, \ and\ \bibinfo {author}
  {\bibfnamefont {R.}~\bibnamefont {Hemley}},\ }\href {\doibase
  10.1103/PhysRevLett.107.255503} {\bibfield  {journal} {\bibinfo  {journal}
  {Phys. Rev. Lett.}\ }\textbf {\bibinfo {volume} {107}},\ \bibinfo {pages}
  {255503} (\bibinfo {year} {2011})}\BibitemShut {NoStop}%
\bibitem [{\citenamefont {Tse}\ \emph {et~al.}(2007)\citenamefont {Tse},
  \citenamefont {Yao},\ and\ \citenamefont {Tanaka}}]{tse_novel_2007}%
  \BibitemOpen
  \bibfield  {author} {\bibinfo {author} {\bibfnamefont {J.~S.}\ \bibnamefont
  {Tse}}, \bibinfo {author} {\bibfnamefont {Y.}~\bibnamefont {Yao}}, \ and\
  \bibinfo {author} {\bibfnamefont {K.}~\bibnamefont {Tanaka}},\ }\href
  {\doibase 10.1103/PhysRevLett.98.117004} {\bibfield  {journal} {\bibinfo
  {journal} {Phy. Rev. Lett.}\ }\textbf {\bibinfo {volume} {98}},\ \bibinfo
  {pages} {117004} (\bibinfo {year} {2007})}\BibitemShut {NoStop}%
\bibitem [{\citenamefont {Chen}\ \emph
  {et~al.}(2008{\natexlab{b}})\citenamefont {Chen}, \citenamefont {Struzhkin},
  \citenamefont {Song}, \citenamefont {Goncharov}, \citenamefont {Ahart},
  \citenamefont {Liu}, \citenamefont {Mao},\ and\ \citenamefont
  {Hemley}}]{Chen_PNAS2008}%
  \BibitemOpen
  \bibfield  {author} {\bibinfo {author} {\bibfnamefont {X.-J.}\ \bibnamefont
  {Chen}}, \bibinfo {author} {\bibfnamefont {V.~V.}\ \bibnamefont {Struzhkin}},
  \bibinfo {author} {\bibfnamefont {Y.}~\bibnamefont {Song}}, \bibinfo {author}
  {\bibfnamefont {A.~F.}\ \bibnamefont {Goncharov}}, \bibinfo {author}
  {\bibfnamefont {M.}~\bibnamefont {Ahart}}, \bibinfo {author} {\bibfnamefont
  {Z.}~\bibnamefont {Liu}}, \bibinfo {author} {\bibfnamefont {H.-k.}\
  \bibnamefont {Mao}}, \ and\ \bibinfo {author} {\bibfnamefont {R.~J.}\
  \bibnamefont {Hemley}},\ }\href {\doibase 10.1073/pnas.0710473105} {\bibfield
   {journal} {\bibinfo  {journal} {Proceedings of the National Academy of
  Sciences}\ }\textbf {\bibinfo {volume} {105}},\ \bibinfo {pages} {20}
  (\bibinfo {year} {2008}{\natexlab{b}})}\BibitemShut {NoStop}%
\bibitem [{\citenamefont {Kim}\ \emph {et~al.}(2008)\citenamefont {Kim},
  \citenamefont {Scheicher}, \citenamefont {Lebègue}, \citenamefont
  {Prasongkit}, \citenamefont {Arnaud}, \citenamefont {Alouani},\ and\
  \citenamefont {Ahuja}}]{Kim_PNAS2008}%
  \BibitemOpen
  \bibfield  {author} {\bibinfo {author} {\bibfnamefont {D.~Y.}\ \bibnamefont
  {Kim}}, \bibinfo {author} {\bibfnamefont {R.~H.}\ \bibnamefont {Scheicher}},
  \bibinfo {author} {\bibfnamefont {S.}~\bibnamefont {Lebègue}}, \bibinfo
  {author} {\bibfnamefont {J.}~\bibnamefont {Prasongkit}}, \bibinfo {author}
  {\bibfnamefont {B.}~\bibnamefont {Arnaud}}, \bibinfo {author} {\bibfnamefont
  {M.}~\bibnamefont {Alouani}}, \ and\ \bibinfo {author} {\bibfnamefont
  {R.}~\bibnamefont {Ahuja}},\ }\href {\doibase 10.1073/pnas.0804148105}
  {\bibfield  {journal} {\bibinfo  {journal} {Proceedings of the National
  Academy of Sciences}\ }\textbf {\bibinfo {volume} {105}},\ \bibinfo {pages}
  {16454} (\bibinfo {year} {2008})}\BibitemShut {NoStop}%
\bibitem [{\citenamefont {Wang}\ \emph {et~al.}(2009)\citenamefont {Wang},
  \citenamefont {Mao}, \citenamefont {Chen},\ and\ \citenamefont
  {Mao}}]{Wang_PNAS2009}%
  \BibitemOpen
  \bibfield  {author} {\bibinfo {author} {\bibfnamefont {S.}~\bibnamefont
  {Wang}}, \bibinfo {author} {\bibfnamefont {H.-k.}\ \bibnamefont {Mao}},
  \bibinfo {author} {\bibfnamefont {X.-J.}\ \bibnamefont {Chen}}, \ and\
  \bibinfo {author} {\bibfnamefont {W.~L.}\ \bibnamefont {Mao}},\ }\href
  {\doibase 10.1073/pnas.0907729106} {\bibfield  {journal} {\bibinfo  {journal}
  {Proceedings of the National Academy of Sciences}\ }\textbf {\bibinfo
  {volume} {106}},\ \bibinfo {pages} {14763} (\bibinfo {year}
  {2009})}\BibitemShut {NoStop}%
\bibitem [{\citenamefont {Yao}\ and\ \citenamefont
  {Klug}(2010)}]{Yao_PNAS2010}%
  \BibitemOpen
  \bibfield  {author} {\bibinfo {author} {\bibfnamefont {Y.}~\bibnamefont
  {Yao}}\ and\ \bibinfo {author} {\bibfnamefont {D.~D.}\ \bibnamefont {Klug}},\
  }\href {\doibase 10.1073/pnas.1006508107} {\bibfield  {journal} {\bibinfo
  {journal} {Proceedings of the National Academy of Sciences}\ }\textbf
  {\bibinfo {volume} {107}},\ \bibinfo {pages} {20893} (\bibinfo {year}
  {2010})}\BibitemShut {NoStop}%
\bibitem [{\citenamefont {Gao}\ \emph {et~al.}(2010)\citenamefont {Gao},
  \citenamefont {Oganov}, \citenamefont {Li}, \citenamefont {Li}, \citenamefont
  {Wang}, \citenamefont {Cui}, \citenamefont {Ma}, \citenamefont {Bergara},
  \citenamefont {Lyakhov}, \citenamefont {Iitaka},\ and\ \citenamefont
  {Zou}}]{gao_high-pressure_2010}%
  \BibitemOpen
  \bibfield  {author} {\bibinfo {author} {\bibfnamefont {G.}~\bibnamefont
  {Gao}}, \bibinfo {author} {\bibfnamefont {A.~R.}\ \bibnamefont {Oganov}},
  \bibinfo {author} {\bibfnamefont {P.}~\bibnamefont {Li}}, \bibinfo {author}
  {\bibfnamefont {Z.}~\bibnamefont {Li}}, \bibinfo {author} {\bibfnamefont
  {H.}~\bibnamefont {Wang}}, \bibinfo {author} {\bibfnamefont {T.}~\bibnamefont
  {Cui}}, \bibinfo {author} {\bibfnamefont {Y.}~\bibnamefont {Ma}}, \bibinfo
  {author} {\bibfnamefont {A.}~\bibnamefont {Bergara}}, \bibinfo {author}
  {\bibfnamefont {A.~O.}\ \bibnamefont {Lyakhov}}, \bibinfo {author}
  {\bibfnamefont {T.}~\bibnamefont {Iitaka}}, \ and\ \bibinfo {author}
  {\bibfnamefont {G.}~\bibnamefont {Zou}},\ }\href {\doibase
  10.1073/pnas.0908342107} {\bibfield  {journal} {\bibinfo  {journal}
  {Proceedings of the National Academy of Sciences}\ }\textbf {\bibinfo
  {volume} {107}},\ \bibinfo {pages} {1317} (\bibinfo {year}
  {2010})}\BibitemShut {NoStop}%
\bibitem [{\citenamefont {Kim}\ \emph {et~al.}(2010)\citenamefont {Kim},
  \citenamefont {Scheicher}, \citenamefont {Mao}, \citenamefont {Kang},\ and\
  \citenamefont {Ahuja}}]{Kim_PNAS2010}%
  \BibitemOpen
  \bibfield  {author} {\bibinfo {author} {\bibfnamefont {D.~Y.}\ \bibnamefont
  {Kim}}, \bibinfo {author} {\bibfnamefont {R.~H.}\ \bibnamefont {Scheicher}},
  \bibinfo {author} {\bibfnamefont {H.-k.}\ \bibnamefont {Mao}}, \bibinfo
  {author} {\bibfnamefont {T.~W.}\ \bibnamefont {Kang}}, \ and\ \bibinfo
  {author} {\bibfnamefont {R.}~\bibnamefont {Ahuja}},\ }\href {\doibase
  10.1073/pnas.0914462107} {\bibfield  {journal} {\bibinfo  {journal}
  {Proceedings of the National Academy of Sciences}\ }\textbf {\bibinfo
  {volume} {107}},\ \bibinfo {pages} {2793} (\bibinfo {year}
  {2010})}\BibitemShut {NoStop}%
\bibitem [{\citenamefont {Li}\ \emph {et~al.}(2010)\citenamefont {Li},
  \citenamefont {Gao}, \citenamefont {Xie}, \citenamefont {Ma}, \citenamefont
  {Cui},\ and\ \citenamefont {Zou}}]{Li_PNAS2010}%
  \BibitemOpen
  \bibfield  {author} {\bibinfo {author} {\bibfnamefont {Y.}~\bibnamefont
  {Li}}, \bibinfo {author} {\bibfnamefont {G.}~\bibnamefont {Gao}}, \bibinfo
  {author} {\bibfnamefont {Y.}~\bibnamefont {Xie}}, \bibinfo {author}
  {\bibfnamefont {Y.}~\bibnamefont {Ma}}, \bibinfo {author} {\bibfnamefont
  {T.}~\bibnamefont {Cui}}, \ and\ \bibinfo {author} {\bibfnamefont
  {G.}~\bibnamefont {Zou}},\ }\href {\doibase 10.1073/pnas.1007354107}
  {\bibfield  {journal} {\bibinfo  {journal} {Proceedings of the National
  Academy of Sciences}\ }\textbf {\bibinfo {volume} {107}},\ \bibinfo {pages}
  {15708} (\bibinfo {year} {2010})}\BibitemShut {NoStop}%
\bibitem [{\citenamefont {Zhou}\ \emph {et~al.}(2012)\citenamefont {Zhou},
  \citenamefont {Jin}, \citenamefont {Meng}, \citenamefont {Bao}, \citenamefont
  {Ma}, \citenamefont {Liu},\ and\ \citenamefont {Cui}}]{Zhou_PRB2012}%
  \BibitemOpen
  \bibfield  {author} {\bibinfo {author} {\bibfnamefont {D.}~\bibnamefont
  {Zhou}}, \bibinfo {author} {\bibfnamefont {X.}~\bibnamefont {Jin}}, \bibinfo
  {author} {\bibfnamefont {X.}~\bibnamefont {Meng}}, \bibinfo {author}
  {\bibfnamefont {G.}~\bibnamefont {Bao}}, \bibinfo {author} {\bibfnamefont
  {Y.}~\bibnamefont {Ma}}, \bibinfo {author} {\bibfnamefont {B.}~\bibnamefont
  {Liu}}, \ and\ \bibinfo {author} {\bibfnamefont {T.}~\bibnamefont {Cui}},\
  }\href {\doibase 10.1103/PhysRevB.86.014118} {\bibfield  {journal} {\bibinfo
  {journal} {Phys. Rev. B}\ }\textbf {\bibinfo {volume} {86}},\ \bibinfo
  {pages} {014118} (\bibinfo {year} {2012})}\BibitemShut {NoStop}%
\bibitem [{\citenamefont {Li}\ \emph {et~al.}(2014)\citenamefont {Li},
  \citenamefont {Hao}, \citenamefont {Liu}, \citenamefont {Li},\ and\
  \citenamefont {Ma}}]{LiYanmingMa_JCP2014}%
  \BibitemOpen
  \bibfield  {author} {\bibinfo {author} {\bibfnamefont {Y.}~\bibnamefont
  {Li}}, \bibinfo {author} {\bibfnamefont {J.}~\bibnamefont {Hao}}, \bibinfo
  {author} {\bibfnamefont {H.}~\bibnamefont {Liu}}, \bibinfo {author}
  {\bibfnamefont {Y.}~\bibnamefont {Li}}, \ and\ \bibinfo {author}
  {\bibfnamefont {Y.}~\bibnamefont {Ma}},\ }\href {\doibase
  http://dx.doi.org/10.1063/1.4874158} {\bibfield  {journal} {\bibinfo
  {journal} {The Journal of Chemical Physics}\ }\textbf {\bibinfo {volume}
  {140}},\  (\bibinfo {year} {2014})}\BibitemShut {NoStop}%
\bibitem [{\citenamefont {Wang}\ and\ \citenamefont
  {Ma}(2014)}]{Yanming_JCP2014}%
  \BibitemOpen
  \bibfield  {author} {\bibinfo {author} {\bibfnamefont {Y.}~\bibnamefont
  {Wang}}\ and\ \bibinfo {author} {\bibfnamefont {Y.}~\bibnamefont {Ma}},\
  }\href {\doibase http://dx.doi.org/10.1063/1.4861966} {\bibfield  {journal}
  {\bibinfo  {journal} {The Journal of Chemical Physics}\ }\textbf {\bibinfo
  {volume} {140}},\ \bibinfo {eid} {040901} (\bibinfo {year}
  {2014})}\BibitemShut {NoStop}%
\bibitem [{\citenamefont {Hooper}\ \emph {et~al.}(2014)\citenamefont {Hooper},
  \citenamefont {Terpstra}, \citenamefont {Shamp},\ and\ \citenamefont
  {Zurek}}]{Hooper_JPC-2014}%
  \BibitemOpen
  \bibfield  {author} {\bibinfo {author} {\bibfnamefont {J.}~\bibnamefont
  {Hooper}}, \bibinfo {author} {\bibfnamefont {T.}~\bibnamefont {Terpstra}},
  \bibinfo {author} {\bibfnamefont {A.}~\bibnamefont {Shamp}}, \ and\ \bibinfo
  {author} {\bibfnamefont {E.}~\bibnamefont {Zurek}},\ }\href {\doibase
  10.1021/jp4125342} {\bibfield  {journal} {\bibinfo  {journal} {The Journal of
  Physical Chemistry C}\ }\textbf {\bibinfo {volume} {118}},\ \bibinfo {pages}
  {6433} (\bibinfo {year} {2014})}\BibitemShut {NoStop}%
\bibitem [{\citenamefont {Duan}\ \emph {et~al.}(2005)\citenamefont {Duan},
  \citenamefont {Liu}, \citenamefont {Tian}, \citenamefont {Li}, \citenamefont
  {Huang}, \citenamefont {Zhao}, \citenamefont {Yu}, \citenamefont {Liu},
  \citenamefont {Tian},\ and\ \citenamefont {Cui}}]{Duan_SciRep2014}%
  \BibitemOpen
  \bibfield  {author} {\bibinfo {author} {\bibfnamefont {D.}~\bibnamefont
  {Duan}}, \bibinfo {author} {\bibfnamefont {Y.}~\bibnamefont {Liu}}, \bibinfo
  {author} {\bibfnamefont {F.}~\bibnamefont {Tian}}, \bibinfo {author}
  {\bibfnamefont {D.}~\bibnamefont {Li}}, \bibinfo {author} {\bibfnamefont
  {X.}~\bibnamefont {Huang}}, \bibinfo {author} {\bibfnamefont
  {Z.}~\bibnamefont {Zhao}}, \bibinfo {author} {\bibfnamefont {H.}~\bibnamefont
  {Yu}}, \bibinfo {author} {\bibfnamefont {B.}~\bibnamefont {Liu}}, \bibinfo
  {author} {\bibfnamefont {W.}~\bibnamefont {Tian}}, \ and\ \bibinfo {author}
  {\bibfnamefont {T.}~\bibnamefont {Cui}},\ }\href {\doibase
  http://dx.doi.org/10.1038/srep06968} {\bibfield  {journal} {\bibinfo
  {journal} {Sci. Rep.}\ }\textbf {\bibinfo {volume} {4}} (\bibinfo {year}
  {2005}),\ http://dx.doi.org/10.1038/srep06968}\BibitemShut {NoStop}%
\bibitem [{\citenamefont {Amsler}\ \emph
  {et~al.}(2012{\natexlab{a}})\citenamefont {Amsler}, \citenamefont
  {Flores-Livas}, \citenamefont {Lehtovaara}, \citenamefont {Balima},
  \citenamefont {Ghasemi}, \citenamefont {Machon}, \citenamefont {Pailh\`es},
  \citenamefont {Willand}, \citenamefont {Caliste}, \citenamefont {Botti},
  \citenamefont {San~Miguel}, \citenamefont {Goedecker},\ and\ \citenamefont
  {Marques}}]{Disilane_JAFL}%
  \BibitemOpen
  \bibfield  {author} {\bibinfo {author} {\bibfnamefont {M.}~\bibnamefont
  {Amsler}}, \bibinfo {author} {\bibfnamefont {J.~A.}\ \bibnamefont
  {Flores-Livas}}, \bibinfo {author} {\bibfnamefont {L.}~\bibnamefont
  {Lehtovaara}}, \bibinfo {author} {\bibfnamefont {F.}~\bibnamefont {Balima}},
  \bibinfo {author} {\bibfnamefont {S.~A.}\ \bibnamefont {Ghasemi}}, \bibinfo
  {author} {\bibfnamefont {D.}~\bibnamefont {Machon}}, \bibinfo {author}
  {\bibfnamefont {S.}~\bibnamefont {Pailh\`es}}, \bibinfo {author}
  {\bibfnamefont {A.}~\bibnamefont {Willand}}, \bibinfo {author} {\bibfnamefont
  {D.}~\bibnamefont {Caliste}}, \bibinfo {author} {\bibfnamefont
  {S.}~\bibnamefont {Botti}}, \bibinfo {author} {\bibfnamefont
  {A.}~\bibnamefont {San~Miguel}}, \bibinfo {author} {\bibfnamefont
  {S.}~\bibnamefont {Goedecker}}, \ and\ \bibinfo {author} {\bibfnamefont
  {M.~A.~L.}\ \bibnamefont {Marques}},\ }\href {\doibase
  10.1103/PhysRevLett.108.065501} {\bibfield  {journal} {\bibinfo  {journal}
  {Phys. Rev. Lett.}\ }\textbf {\bibinfo {volume} {108}},\ \bibinfo {pages}
  {065501} (\bibinfo {year} {2012}{\natexlab{a}})}\BibitemShut {NoStop}%
\bibitem [{\citenamefont {{Bernstein}}\ \emph {et~al.}(2015)\citenamefont
  {{Bernstein}}, \citenamefont {{Hellberg}}, \citenamefont {{Johannes}},
  \citenamefont {{Mazin}},\ and\ \citenamefont
  {{Mehl}}}]{Bernstein_arxiv_2014}%
  \BibitemOpen
  \bibfield  {author} {\bibinfo {author} {\bibfnamefont {N.}~\bibnamefont
  {{Bernstein}}}, \bibinfo {author} {\bibfnamefont {C.~S.}\ \bibnamefont
  {{Hellberg}}}, \bibinfo {author} {\bibfnamefont {M.~D.}\ \bibnamefont
  {{Johannes}}}, \bibinfo {author} {\bibfnamefont {I.~I.}\ \bibnamefont
  {{Mazin}}}, \ and\ \bibinfo {author} {\bibfnamefont {M.~J.}\ \bibnamefont
  {{Mehl}}},\ }\href@noop {} {\bibfield  {journal} {\bibinfo  {journal} {ArXiv
  e-prints}\ } (\bibinfo {year} {2015})},\ \Eprint
  {http://arxiv.org/abs/1501.00196} {arXiv:1501.00196 [cond-mat.supr-con]}
  \BibitemShut {NoStop}%
\bibitem [{\citenamefont {Goedecker}(2004)}]{Goedecker_2004}%
  \BibitemOpen
  \bibfield  {author} {\bibinfo {author} {\bibfnamefont {S.}~\bibnamefont
  {Goedecker}},\ }\href@noop {} {\bibfield  {journal} {\bibinfo  {journal} {The
  Journal of Chemical Physics}\ }\textbf {\bibinfo {volume} {120}},\ \bibinfo
  {pages} {9911} (\bibinfo {year} {2004})}\BibitemShut {NoStop}%
\bibitem [{\citenamefont {Goedecker}\ \emph {et~al.}(2005)\citenamefont
  {Goedecker}, \citenamefont {Hellmann},\ and\ \citenamefont
  {Lenosky}}]{Goedecker_2005}%
  \BibitemOpen
  \bibfield  {author} {\bibinfo {author} {\bibfnamefont {S.}~\bibnamefont
  {Goedecker}}, \bibinfo {author} {\bibfnamefont {W.}~\bibnamefont {Hellmann}},
  \ and\ \bibinfo {author} {\bibfnamefont {T.}~\bibnamefont {Lenosky}},\
  }\href@noop {} {\bibfield  {journal} {\bibinfo  {journal} {Phys. Rev. Lett.}\
  }\textbf {\bibinfo {volume} {95}},\ \bibinfo {pages} {055501} (\bibinfo
  {year} {2005})}\BibitemShut {NoStop}%
\bibitem [{\citenamefont {Amsler}\ and\ \citenamefont
  {Goedecker}(2010)}]{Amsler_2010}%
  \BibitemOpen
  \bibfield  {author} {\bibinfo {author} {\bibfnamefont {M.}~\bibnamefont
  {Amsler}}\ and\ \bibinfo {author} {\bibfnamefont {S.}~\bibnamefont
  {Goedecker}},\ }\href@noop {} {\bibfield  {journal} {\bibinfo  {journal} {The
  Journal of Chemical Physics}\ }\textbf {\bibinfo {volume} {133}},\ \bibinfo
  {pages} {224104} (\bibinfo {year} {2010})}\BibitemShut {NoStop}%
\bibitem [{\citenamefont {Gonze}\ \emph {et~al.}(2009)\citenamefont {Gonze},
  \citenamefont {Amadon}, \citenamefont {Anglade}, \citenamefont {Beuken},
  \citenamefont {Bottin}, \citenamefont {Boulanger}, \citenamefont {Bruneval},
  \citenamefont {Caliste}, \citenamefont {Caracas}, \citenamefont
  {C\^{o}t\'{e}}, \citenamefont {Deutsch}, \citenamefont {Genovese},
  \citenamefont {Ghosez}, \citenamefont {Giantomassi}, \citenamefont
  {Goedecker}, \citenamefont {Hamann}, \citenamefont {Hermet}, \citenamefont
  {Jollet}, \citenamefont {Jomard}, \citenamefont {Leroux}, \citenamefont
  {Mancini}, \citenamefont {Mazevet}, \citenamefont {Oliveira}, \citenamefont
  {Onida}, \citenamefont {Pouillon}, \citenamefont {Rangel}, \citenamefont
  {Rignanese}, \citenamefont {Sangalli}, \citenamefont {Shaltaf}, \citenamefont
  {Torrent}, \citenamefont {Verstraete}, \citenamefont {Zerah},\ and\
  \citenamefont {Zwanziger}}]{gonze_abinit_2009}%
  \BibitemOpen
  \bibfield  {author} {\bibinfo {author} {\bibfnamefont {X.}~\bibnamefont
  {Gonze}}, \bibinfo {author} {\bibfnamefont {B.}~\bibnamefont {Amadon}},
  \bibinfo {author} {\bibfnamefont {P.}~\bibnamefont {Anglade}}, \bibinfo
  {author} {\bibfnamefont {J.}~\bibnamefont {Beuken}}, \bibinfo {author}
  {\bibfnamefont {F.}~\bibnamefont {Bottin}}, \bibinfo {author} {\bibfnamefont
  {P.}~\bibnamefont {Boulanger}}, \bibinfo {author} {\bibfnamefont
  {F.}~\bibnamefont {Bruneval}}, \bibinfo {author} {\bibfnamefont
  {D.}~\bibnamefont {Caliste}}, \bibinfo {author} {\bibfnamefont
  {R.}~\bibnamefont {Caracas}}, \bibinfo {author} {\bibfnamefont
  {M.}~\bibnamefont {C\^{o}t\'{e}}}, \bibinfo {author} {\bibfnamefont
  {T.}~\bibnamefont {Deutsch}}, \bibinfo {author} {\bibfnamefont
  {L.}~\bibnamefont {Genovese}}, \bibinfo {author} {\bibfnamefont
  {P.}~\bibnamefont {Ghosez}}, \bibinfo {author} {\bibfnamefont
  {M.}~\bibnamefont {Giantomassi}}, \bibinfo {author} {\bibfnamefont
  {S.}~\bibnamefont {Goedecker}}, \bibinfo {author} {\bibfnamefont
  {D.}~\bibnamefont {Hamann}}, \bibinfo {author} {\bibfnamefont
  {P.}~\bibnamefont {Hermet}}, \bibinfo {author} {\bibfnamefont
  {F.}~\bibnamefont {Jollet}}, \bibinfo {author} {\bibfnamefont
  {G.}~\bibnamefont {Jomard}}, \bibinfo {author} {\bibfnamefont
  {S.}~\bibnamefont {Leroux}}, \bibinfo {author} {\bibfnamefont
  {M.}~\bibnamefont {Mancini}}, \bibinfo {author} {\bibfnamefont
  {S.}~\bibnamefont {Mazevet}}, \bibinfo {author} {\bibfnamefont
  {M.}~\bibnamefont {Oliveira}}, \bibinfo {author} {\bibfnamefont
  {G.}~\bibnamefont {Onida}}, \bibinfo {author} {\bibfnamefont
  {Y.}~\bibnamefont {Pouillon}}, \bibinfo {author} {\bibfnamefont
  {T.}~\bibnamefont {Rangel}}, \bibinfo {author} {\bibfnamefont
  {G.}~\bibnamefont {Rignanese}}, \bibinfo {author} {\bibfnamefont
  {D.}~\bibnamefont {Sangalli}}, \bibinfo {author} {\bibfnamefont
  {R.}~\bibnamefont {Shaltaf}}, \bibinfo {author} {\bibfnamefont
  {M.}~\bibnamefont {Torrent}}, \bibinfo {author} {\bibfnamefont
  {M.}~\bibnamefont {Verstraete}}, \bibinfo {author} {\bibfnamefont
  {G.}~\bibnamefont {Zerah}}, \ and\ \bibinfo {author} {\bibfnamefont
  {J.}~\bibnamefont {Zwanziger}},\ }\href@noop {} {\bibfield  {journal}
  {\bibinfo  {journal} {Computer Physics Communications}\ }\textbf {\bibinfo
  {volume} {180}},\ \bibinfo {pages} {2582} (\bibinfo {year}
  {2009})}\BibitemShut {NoStop}%
\bibitem [{\citenamefont {Giannozzi}\ \emph {et~al.}(2009)\citenamefont
  {Giannozzi}, \citenamefont {Baroni}, \citenamefont {Bonini}, \citenamefont
  {Calandra}, \citenamefont {Car}, \citenamefont {Cavazzoni}, \citenamefont
  {Ceresoli}, \citenamefont {Chiarotti}, \citenamefont {Cococcioni},
  \citenamefont {Dabo}, \citenamefont {{Dal Corso}}, \citenamefont
  {de~Gironcoli}, \citenamefont {Fabris}, \citenamefont {Fratesi},
  \citenamefont {Gebauer}, \citenamefont {Gerstmann}, \citenamefont
  {Gougoussis}, \citenamefont {Kokalj}, \citenamefont {Lazzeri}, \citenamefont
  {Martin-Samos}, \citenamefont {Marzari}, \citenamefont {Mauri}, \citenamefont
  {Mazzarello}, \citenamefont {Paolini}, \citenamefont {Pasquarello},
  \citenamefont {Paulatto}, \citenamefont {Sbraccia}, \citenamefont {Scandolo},
  \citenamefont {Sclauzero}, \citenamefont {Seitsonen}, \citenamefont
  {Smogunov}, \citenamefont {Umari},\ and\ \citenamefont
  {Wentzcovitch}}]{QE-2009}%
  \BibitemOpen
  \bibfield  {author} {\bibinfo {author} {\bibfnamefont {P.}~\bibnamefont
  {Giannozzi}}, \bibinfo {author} {\bibfnamefont {S.}~\bibnamefont {Baroni}},
  \bibinfo {author} {\bibfnamefont {N.}~\bibnamefont {Bonini}}, \bibinfo
  {author} {\bibfnamefont {M.}~\bibnamefont {Calandra}}, \bibinfo {author}
  {\bibfnamefont {R.}~\bibnamefont {Car}}, \bibinfo {author} {\bibfnamefont
  {C.}~\bibnamefont {Cavazzoni}}, \bibinfo {author} {\bibfnamefont
  {D.}~\bibnamefont {Ceresoli}}, \bibinfo {author} {\bibfnamefont {G.~L.}\
  \bibnamefont {Chiarotti}}, \bibinfo {author} {\bibfnamefont {M.}~\bibnamefont
  {Cococcioni}}, \bibinfo {author} {\bibfnamefont {I.}~\bibnamefont {Dabo}},
  \bibinfo {author} {\bibfnamefont {A.}~\bibnamefont {{Dal Corso}}}, \bibinfo
  {author} {\bibfnamefont {S.}~\bibnamefont {de~Gironcoli}}, \bibinfo {author}
  {\bibfnamefont {S.}~\bibnamefont {Fabris}}, \bibinfo {author} {\bibfnamefont
  {G.}~\bibnamefont {Fratesi}}, \bibinfo {author} {\bibfnamefont
  {R.}~\bibnamefont {Gebauer}}, \bibinfo {author} {\bibfnamefont
  {U.}~\bibnamefont {Gerstmann}}, \bibinfo {author} {\bibfnamefont
  {C.}~\bibnamefont {Gougoussis}}, \bibinfo {author} {\bibfnamefont
  {A.}~\bibnamefont {Kokalj}}, \bibinfo {author} {\bibfnamefont
  {M.}~\bibnamefont {Lazzeri}}, \bibinfo {author} {\bibfnamefont
  {L.}~\bibnamefont {Martin-Samos}}, \bibinfo {author} {\bibfnamefont
  {N.}~\bibnamefont {Marzari}}, \bibinfo {author} {\bibfnamefont
  {F.}~\bibnamefont {Mauri}}, \bibinfo {author} {\bibfnamefont
  {R.}~\bibnamefont {Mazzarello}}, \bibinfo {author} {\bibfnamefont
  {S.}~\bibnamefont {Paolini}}, \bibinfo {author} {\bibfnamefont
  {A.}~\bibnamefont {Pasquarello}}, \bibinfo {author} {\bibfnamefont
  {L.}~\bibnamefont {Paulatto}}, \bibinfo {author} {\bibfnamefont
  {C.}~\bibnamefont {Sbraccia}}, \bibinfo {author} {\bibfnamefont
  {S.}~\bibnamefont {Scandolo}}, \bibinfo {author} {\bibfnamefont
  {G.}~\bibnamefont {Sclauzero}}, \bibinfo {author} {\bibfnamefont {A.~P.}\
  \bibnamefont {Seitsonen}}, \bibinfo {author} {\bibfnamefont {A.}~\bibnamefont
  {Smogunov}}, \bibinfo {author} {\bibfnamefont {P.}~\bibnamefont {Umari}}, \
  and\ \bibinfo {author} {\bibfnamefont {R.~M.}\ \bibnamefont {Wentzcovitch}},\
  }\href {http://www.quantum-espresso.org} {\bibfield  {journal} {\bibinfo
  {journal} {Journal of Physics: Condensed Matter}\ }\textbf {\bibinfo {volume}
  {21}},\ \bibinfo {pages} {395502 (19pp)} (\bibinfo {year}
  {2009})}\BibitemShut {NoStop}%
\bibitem [{\citenamefont {Fuchs}\ and\ \citenamefont
  {Scheffler}(1999)}]{FHI_Fuchs}%
  \BibitemOpen
  \bibfield  {author} {\bibinfo {author} {\bibfnamefont {M.}~\bibnamefont
  {Fuchs}}\ and\ \bibinfo {author} {\bibfnamefont {M.}~\bibnamefont
  {Scheffler}},\ }\href@noop {} {\bibfield  {journal} {\bibinfo  {journal}
  {Comput. Phys. Commun.}\ }\textbf {\bibinfo {volume} {119}},\ \bibinfo
  {pages} {67} (\bibinfo {year} {1999})}\BibitemShut {NoStop}%
\bibitem [{\citenamefont {Amsler}\ \emph
  {et~al.}(2012{\natexlab{b}})\citenamefont {Amsler}, \citenamefont
  {Flores-Livas}, \citenamefont {Lehtovaara}, \citenamefont {Balima},
  \citenamefont {Ghasemi}, \citenamefont {Machon}, \citenamefont {Pailh\`es},
  \citenamefont {Willand}, \citenamefont {Caliste}, \citenamefont {Botti},
  \citenamefont {San~Miguel}, \citenamefont {Goedecker},\ and\ \citenamefont
  {Marques}}]{MA_JAFL}%
  \BibitemOpen
  \bibfield  {author} {\bibinfo {author} {\bibfnamefont {M.}~\bibnamefont
  {Amsler}}, \bibinfo {author} {\bibfnamefont {J.~A.}\ \bibnamefont
  {Flores-Livas}}, \bibinfo {author} {\bibfnamefont {L.}~\bibnamefont
  {Lehtovaara}}, \bibinfo {author} {\bibfnamefont {F.}~\bibnamefont {Balima}},
  \bibinfo {author} {\bibfnamefont {S.~A.}\ \bibnamefont {Ghasemi}}, \bibinfo
  {author} {\bibfnamefont {D.}~\bibnamefont {Machon}}, \bibinfo {author}
  {\bibfnamefont {S.}~\bibnamefont {Pailh\`es}}, \bibinfo {author}
  {\bibfnamefont {A.}~\bibnamefont {Willand}}, \bibinfo {author} {\bibfnamefont
  {D.}~\bibnamefont {Caliste}}, \bibinfo {author} {\bibfnamefont
  {S.}~\bibnamefont {Botti}}, \bibinfo {author} {\bibfnamefont
  {A.}~\bibnamefont {San~Miguel}}, \bibinfo {author} {\bibfnamefont
  {S.}~\bibnamefont {Goedecker}}, \ and\ \bibinfo {author} {\bibfnamefont
  {M.~A.~L.}\ \bibnamefont {Marques}},\ }\href {\doibase
  10.1103/PhysRevLett.108.065501} {\bibfield  {journal} {\bibinfo  {journal}
  {Phys. Rev. Lett.}\ }\textbf {\bibinfo {volume} {108}},\ \bibinfo {pages}
  {065501} (\bibinfo {year} {2012}{\natexlab{b}})}\BibitemShut {NoStop}%
\bibitem [{\citenamefont {Amsler}\ \emph
  {et~al.}(2012{\natexlab{c}})\citenamefont {Amsler}, \citenamefont
  {Flores-Livas}, \citenamefont {Huan}, \citenamefont {Botti}, \citenamefont
  {Marques},\ and\ \citenamefont {Goedecker}}]{LiAlH_Maxmotif}%
  \BibitemOpen
  \bibfield  {author} {\bibinfo {author} {\bibfnamefont {M.}~\bibnamefont
  {Amsler}}, \bibinfo {author} {\bibfnamefont {J.~A.}\ \bibnamefont
  {Flores-Livas}}, \bibinfo {author} {\bibfnamefont {T.~D.}\ \bibnamefont
  {Huan}}, \bibinfo {author} {\bibfnamefont {S.}~\bibnamefont {Botti}},
  \bibinfo {author} {\bibfnamefont {M.~A.~L.}\ \bibnamefont {Marques}}, \ and\
  \bibinfo {author} {\bibfnamefont {S.}~\bibnamefont {Goedecker}},\ }\href
  {\doibase 10.1103/PhysRevLett.108.205505} {\bibfield  {journal} {\bibinfo
  {journal} {Phys. Rev. Lett.}\ }\textbf {\bibinfo {volume} {108}},\ \bibinfo
  {pages} {205505} (\bibinfo {year} {2012}{\natexlab{c}})}\BibitemShut
  {NoStop}%
\bibitem [{\citenamefont {Botti}\ \emph {et~al.}(2012)\citenamefont {Botti},
  \citenamefont {Flores-Livas}, \citenamefont {Amsler}, \citenamefont
  {Goedecker},\ and\ \citenamefont {Marques}}]{BJAFL_PRB2012}%
  \BibitemOpen
  \bibfield  {author} {\bibinfo {author} {\bibfnamefont {S.}~\bibnamefont
  {Botti}}, \bibinfo {author} {\bibfnamefont {J.~A.}\ \bibnamefont
  {Flores-Livas}}, \bibinfo {author} {\bibfnamefont {M.}~\bibnamefont
  {Amsler}}, \bibinfo {author} {\bibfnamefont {S.}~\bibnamefont {Goedecker}}, \
  and\ \bibinfo {author} {\bibfnamefont {M.~A.~L.}\ \bibnamefont {Marques}},\
  }\href {\doibase 10.1103/PhysRevB.86.121204} {\bibfield  {journal} {\bibinfo
  {journal} {Phys. Rev. B}\ }\textbf {\bibinfo {volume} {86}},\ \bibinfo
  {pages} {121204} (\bibinfo {year} {2012})}\BibitemShut {NoStop}%
\bibitem [{\citenamefont {Bitzek}\ \emph {et~al.}(2006)\citenamefont {Bitzek},
  \citenamefont {Koskinen}, \citenamefont {G\"ahler}, \citenamefont {Moseler},\
  and\ \citenamefont {Gumbsch}}]{FIRE_2006}%
  \BibitemOpen
  \bibfield  {author} {\bibinfo {author} {\bibfnamefont {E.}~\bibnamefont
  {Bitzek}}, \bibinfo {author} {\bibfnamefont {P.}~\bibnamefont {Koskinen}},
  \bibinfo {author} {\bibfnamefont {F.}~\bibnamefont {G\"ahler}}, \bibinfo
  {author} {\bibfnamefont {M.}~\bibnamefont {Moseler}}, \ and\ \bibinfo
  {author} {\bibfnamefont {P.}~\bibnamefont {Gumbsch}},\ }\href {\doibase
  10.1103/PhysRevLett.97.170201} {\bibfield  {journal} {\bibinfo  {journal}
  {Phys. Rev. Lett.}\ }\textbf {\bibinfo {volume} {97}},\ \bibinfo {pages}
  {170201} (\bibinfo {year} {2006})}\BibitemShut {NoStop}%
\bibitem [{\citenamefont {Kresse}\ and\ \citenamefont
  {Furthm\"{u}ller}(1996)}]{VASP_Kresse}%
  \BibitemOpen
  \bibfield  {author} {\bibinfo {author} {\bibfnamefont {G.}~\bibnamefont
  {Kresse}}\ and\ \bibinfo {author} {\bibfnamefont {J.}~\bibnamefont
  {Furthm\"{u}ller}},\ }\href@noop {} {\bibfield  {journal} {\bibinfo
  {journal} {Comput. Mat. Sci.}\ }\textbf {\bibinfo {volume} {6}},\ \bibinfo
  {pages} {15} (\bibinfo {year} {1996})}\BibitemShut {NoStop}%
\bibitem [{\citenamefont {Oliveira}\ \emph {et~al.}(1988)\citenamefont
  {Oliveira}, \citenamefont {Gross},\ and\ \citenamefont
  {Kohn}}]{OGK_SCDFT_PRL1988}%
  \BibitemOpen
  \bibfield  {author} {\bibinfo {author} {\bibfnamefont {L.~N.}\ \bibnamefont
  {Oliveira}}, \bibinfo {author} {\bibfnamefont {E.~K.~U.}\ \bibnamefont
  {Gross}}, \ and\ \bibinfo {author} {\bibfnamefont {W.}~\bibnamefont {Kohn}},\
  }\href {\doibase 10.1103/PhysRevLett.60.2430} {\bibfield  {journal} {\bibinfo
   {journal} {Phys. Rev. Lett.}\ }\textbf {\bibinfo {volume} {60}},\ \bibinfo
  {pages} {2430} (\bibinfo {year} {1988})}\BibitemShut {NoStop}%
\bibitem [{\citenamefont {L\"uders}\ \emph {et~al.}(2005)\citenamefont
  {L\"uders}, \citenamefont {Marques}, \citenamefont {Lathiotakis},
  \citenamefont {Floris}, \citenamefont {Profeta}, \citenamefont {Fast},
  \citenamefont {Continenza}, \citenamefont {Massidda},\ and\ \citenamefont
  {Gross}}]{Lueders_SCDFT_PRB2005}%
  \BibitemOpen
  \bibfield  {author} {\bibinfo {author} {\bibfnamefont {M.}~\bibnamefont
  {L\"uders}}, \bibinfo {author} {\bibfnamefont {M.~A.~L.}\ \bibnamefont
  {Marques}}, \bibinfo {author} {\bibfnamefont {N.~N.}\ \bibnamefont
  {Lathiotakis}}, \bibinfo {author} {\bibfnamefont {A.}~\bibnamefont {Floris}},
  \bibinfo {author} {\bibfnamefont {G.}~\bibnamefont {Profeta}}, \bibinfo
  {author} {\bibfnamefont {L.}~\bibnamefont {Fast}}, \bibinfo {author}
  {\bibfnamefont {A.}~\bibnamefont {Continenza}}, \bibinfo {author}
  {\bibfnamefont {S.}~\bibnamefont {Massidda}}, \ and\ \bibinfo {author}
  {\bibfnamefont {E.~K.~U.}\ \bibnamefont {Gross}},\ }\href {\doibase
  10.1103/PhysRevB.72.024545} {\bibfield  {journal} {\bibinfo  {journal} {Phys.
  Rev. B}\ }\textbf {\bibinfo {volume} {72}},\ \bibinfo {pages} {024545}
  (\bibinfo {year} {2005})}\BibitemShut {NoStop}%
\bibitem [{\citenamefont {Marques}\ \emph {et~al.}(2005)\citenamefont
  {Marques}, \citenamefont {L\"uders}, \citenamefont {Lathiotakis},
  \citenamefont {Profeta}, \citenamefont {Floris}, \citenamefont {Fast},
  \citenamefont {Continenza}, \citenamefont {Gross},\ and\ \citenamefont
  {Massidda}}]{Marques_SCDFT_PRB2005}%
  \BibitemOpen
  \bibfield  {author} {\bibinfo {author} {\bibfnamefont {M.~A.~L.}\
  \bibnamefont {Marques}}, \bibinfo {author} {\bibfnamefont {M.}~\bibnamefont
  {L\"uders}}, \bibinfo {author} {\bibfnamefont {N.~N.}\ \bibnamefont
  {Lathiotakis}}, \bibinfo {author} {\bibfnamefont {G.}~\bibnamefont
  {Profeta}}, \bibinfo {author} {\bibfnamefont {A.}~\bibnamefont {Floris}},
  \bibinfo {author} {\bibfnamefont {L.}~\bibnamefont {Fast}}, \bibinfo {author}
  {\bibfnamefont {A.}~\bibnamefont {Continenza}}, \bibinfo {author}
  {\bibfnamefont {E.~K.~U.}\ \bibnamefont {Gross}}, \ and\ \bibinfo {author}
  {\bibfnamefont {S.}~\bibnamefont {Massidda}},\ }\href {\doibase
  10.1103/PhysRevB.72.024546} {\bibfield  {journal} {\bibinfo  {journal} {Phys.
  Rev. B}\ }\textbf {\bibinfo {volume} {72}},\ \bibinfo {pages} {024546}
  (\bibinfo {year} {2005})}\BibitemShut {NoStop}%
\bibitem [{\citenamefont {Floris}\ \emph {et~al.}(2007)\citenamefont {Floris},
  \citenamefont {Sanna}, \citenamefont {Massidda},\ and\ \citenamefont
  {Gross}}]{Floris_Pb_PRB2007}%
  \BibitemOpen
  \bibfield  {author} {\bibinfo {author} {\bibfnamefont {A.}~\bibnamefont
  {Floris}}, \bibinfo {author} {\bibfnamefont {A.}~\bibnamefont {Sanna}},
  \bibinfo {author} {\bibfnamefont {S.}~\bibnamefont {Massidda}}, \ and\
  \bibinfo {author} {\bibfnamefont {E.~K.~U.}\ \bibnamefont {Gross}},\ }\href
  {\doibase 10.1103/PhysRevB.75.054508} {\bibfield  {journal} {\bibinfo
  {journal} {Phys. Rev. B}\ }\textbf {\bibinfo {volume} {75}},\ \bibinfo
  {pages} {054508} (\bibinfo {year} {2007})}\BibitemShut {NoStop}%
\bibitem [{\citenamefont {Gonnelli}\ \emph {et~al.}(2008)\citenamefont
  {Gonnelli}, \citenamefont {Daghero}, \citenamefont {Delaude}, \citenamefont
  {Tortello}, \citenamefont {Ummarino}, \citenamefont {Stepanov}, \citenamefont
  {Kim}, \citenamefont {Kremer}, \citenamefont {Sanna}, \citenamefont
  {Profeta},\ and\ \citenamefont {Massidda}}]{Gonnelli_CaC6_PRL2008}%
  \BibitemOpen
  \bibfield  {author} {\bibinfo {author} {\bibfnamefont {R.~S.}\ \bibnamefont
  {Gonnelli}}, \bibinfo {author} {\bibfnamefont {D.}~\bibnamefont {Daghero}},
  \bibinfo {author} {\bibfnamefont {D.}~\bibnamefont {Delaude}}, \bibinfo
  {author} {\bibfnamefont {M.}~\bibnamefont {Tortello}}, \bibinfo {author}
  {\bibfnamefont {G.~A.}\ \bibnamefont {Ummarino}}, \bibinfo {author}
  {\bibfnamefont {V.~A.}\ \bibnamefont {Stepanov}}, \bibinfo {author}
  {\bibfnamefont {J.~S.}\ \bibnamefont {Kim}}, \bibinfo {author} {\bibfnamefont
  {R.~K.}\ \bibnamefont {Kremer}}, \bibinfo {author} {\bibfnamefont
  {A.}~\bibnamefont {Sanna}}, \bibinfo {author} {\bibfnamefont
  {G.}~\bibnamefont {Profeta}}, \ and\ \bibinfo {author} {\bibfnamefont
  {S.}~\bibnamefont {Massidda}},\ }\href {\doibase
  10.1103/PhysRevLett.100.207004} {\bibfield  {journal} {\bibinfo  {journal}
  {Phys. Rev. Lett.}\ }\textbf {\bibinfo {volume} {100}},\ \bibinfo {pages}
  {207004} (\bibinfo {year} {2008})}\BibitemShut {NoStop}%
\bibitem [{\citenamefont {Sanna}\ \emph {et~al.}(2007)\citenamefont {Sanna},
  \citenamefont {Profeta}, \citenamefont {Floris}, \citenamefont {Marini},
  \citenamefont {Gross},\ and\ \citenamefont
  {Massidda}}]{Sanna_CaC6_RapCom2007}%
  \BibitemOpen
  \bibfield  {author} {\bibinfo {author} {\bibfnamefont {A.}~\bibnamefont
  {Sanna}}, \bibinfo {author} {\bibfnamefont {G.}~\bibnamefont {Profeta}},
  \bibinfo {author} {\bibfnamefont {A.}~\bibnamefont {Floris}}, \bibinfo
  {author} {\bibfnamefont {A.}~\bibnamefont {Marini}}, \bibinfo {author}
  {\bibfnamefont {E.~K.~U.}\ \bibnamefont {Gross}}, \ and\ \bibinfo {author}
  {\bibfnamefont {S.}~\bibnamefont {Massidda}},\ }\href {\doibase
  10.1103/PhysRevB.75.020511} {\bibfield  {journal} {\bibinfo  {journal} {Phys.
  Rev. B}\ }\textbf {\bibinfo {volume} {75}},\ \bibinfo {pages} {020511}
  (\bibinfo {year} {2007})}\BibitemShut {NoStop}%
\bibitem [{\citenamefont {Profeta}\ \emph {et~al.}(2006)\citenamefont
  {Profeta}, \citenamefont {Franchini}, \citenamefont {Lathiotakis},
  \citenamefont {Floris}, \citenamefont {Sanna}, \citenamefont {Marques},
  \citenamefont {L\"uders}, \citenamefont {Massidda}, \citenamefont {Gross},\
  and\ \citenamefont {Continenza}}]{Profeta_LiKAl_PRL2006}%
  \BibitemOpen
  \bibfield  {author} {\bibinfo {author} {\bibfnamefont {G.}~\bibnamefont
  {Profeta}}, \bibinfo {author} {\bibfnamefont {C.}~\bibnamefont {Franchini}},
  \bibinfo {author} {\bibfnamefont {N.}~\bibnamefont {Lathiotakis}}, \bibinfo
  {author} {\bibfnamefont {A.}~\bibnamefont {Floris}}, \bibinfo {author}
  {\bibfnamefont {A.}~\bibnamefont {Sanna}}, \bibinfo {author} {\bibfnamefont
  {M.~A.~L.}\ \bibnamefont {Marques}}, \bibinfo {author} {\bibfnamefont
  {M.}~\bibnamefont {L\"uders}}, \bibinfo {author} {\bibfnamefont
  {S.}~\bibnamefont {Massidda}}, \bibinfo {author} {\bibfnamefont {E.~K.~U.}\
  \bibnamefont {Gross}}, \ and\ \bibinfo {author} {\bibfnamefont
  {A.}~\bibnamefont {Continenza}},\ }\href {\doibase
  10.1103/PhysRevLett.96.047003} {\bibfield  {journal} {\bibinfo  {journal}
  {Phys. Rev. Lett.}\ }\textbf {\bibinfo {volume} {96}},\ \bibinfo {pages}
  {047003} (\bibinfo {year} {2006})}\BibitemShut {NoStop}%
\bibitem [{Note1()}]{Note1}%
  \BibitemOpen
  \bibinfo {note} {The phononic functional we use is an improved version with
  respect to Ref.~\cite {Lueders_SCDFT_PRB2005,Marques_SCDFT_PRB2005} and is
  discussed in Ref.~\cite {Sanna_Migdal}. In this work Coulomb interactions are
  included within static RPA~\cite {Sanna_CaC6_RapCom2007}, therefore excluding
  magnetic source of coupling\cite {Frank_SF_PRB2014}}\BibitemShut {NoStop}%
\bibitem [{Note2()}]{Note2}%
  \BibitemOpen
  \bibinfo {note} {The decomposition enthalpies have been computed from the
  predicted structures of hydrogen $P6_3m$ and $C2/c$~\cite
  {pickard_structure_2007} and for sulfur and selenium on the $R3m$ and
  $Im$\={3}$m$ reported to occur at high pressure.~\cite
  {Akahama_S_PRB1993,Akahama_Se_PRB1993,Akahama_Se_met_PRB1997,Shimizu_EHP2005}}\BibitemShut
  {NoStop}%
\bibitem [{\citenamefont {Carbotte}(1990)}]{Carbotte_RMP1990}%
  \BibitemOpen
  \bibfield  {author} {\bibinfo {author} {\bibfnamefont {J.~P.}\ \bibnamefont
  {Carbotte}},\ }\href {\doibase 10.1103/RevModPhys.62.1027} {\bibfield
  {journal} {\bibinfo  {journal} {Rev. Mod. Phys.}\ }\textbf {\bibinfo {volume}
  {62}},\ \bibinfo {pages} {1027} (\bibinfo {year} {1990})}\BibitemShut
  {NoStop}%
\bibitem [{\citenamefont {Allen}\ and\ \citenamefont {Mitrovi{\'
  c}}(1983)}]{AllenMitrovic1983}%
  \BibitemOpen
  \bibfield  {author} {\bibinfo {author} {\bibfnamefont {P.~B.}\ \bibnamefont
  {Allen}}\ and\ \bibinfo {author} {\bibfnamefont {B.}~\bibnamefont {Mitrovi{\'
  c}}}\ }(\bibinfo  {publisher} {Academic Press},\ \bibinfo {year} {1983})\
  pp.\ \bibinfo {pages} {1 -- 92}\BibitemShut {NoStop}%
\bibitem [{\citenamefont {{Papaconstantopoulos}}\ \emph
  {et~al.}(2015)\citenamefont {{Papaconstantopoulos}}, \citenamefont {{Klein}},
  \citenamefont {{Mehl}},\ and\ \citenamefont {{Pickett}}}]{Pickett_Arxiv2015}%
  \BibitemOpen
  \bibfield  {author} {\bibinfo {author} {\bibfnamefont {D.~A.}\ \bibnamefont
  {{Papaconstantopoulos}}}, \bibinfo {author} {\bibfnamefont {B.~M.}\
  \bibnamefont {{Klein}}}, \bibinfo {author} {\bibfnamefont {M.~J.}\
  \bibnamefont {{Mehl}}}, \ and\ \bibinfo {author} {\bibfnamefont {W.~E.}\
  \bibnamefont {{Pickett}}},\ }\href@noop {} {\bibfield  {journal} {\bibinfo
  {journal} {ArXiv e-prints}\ } (\bibinfo {year} {2015})},\ \Eprint
  {http://arxiv.org/abs/1501.03950} {arXiv:1501.03950 [cond-mat.supr-con]}
  \BibitemShut {NoStop}%
\bibitem [{Note3()}]{Note3}%
  \BibitemOpen
  \bibinfo {note} {Conventional implementation of the Eliashberg equations due
  to their computational cost, usually assume a $k$-independent pairing and a
  flat density of states. Anisotropic implementations \cite
  {Margine_anisoEliashberg_PRB2013} are not intrinsically affected by this
  limit.}\BibitemShut {Stop}%
\bibitem [{\citenamefont {Allen}\ and\ \citenamefont
  {Dynes}(1975)}]{AllenDynes_PRB1975}%
  \BibitemOpen
  \bibfield  {author} {\bibinfo {author} {\bibfnamefont {P.~B.}\ \bibnamefont
  {Allen}}\ and\ \bibinfo {author} {\bibfnamefont {R.~C.}\ \bibnamefont
  {Dynes}},\ }\href {\doibase 10.1103/PhysRevB.12.905} {\bibfield  {journal}
  {\bibinfo  {journal} {Phys. Rev. B}\ }\textbf {\bibinfo {volume} {12}},\
  \bibinfo {pages} {905} (\bibinfo {year} {1975})}\BibitemShut {NoStop}%
\bibitem [{\citenamefont {McMillan}(1968)}]{McMillanTc}%
  \BibitemOpen
  \bibfield  {author} {\bibinfo {author} {\bibfnamefont {W.~L.}\ \bibnamefont
  {McMillan}},\ }\href {\doibase 10.1103/PhysRev.167.331} {\bibfield  {journal}
  {\bibinfo  {journal} {Phys. Rev.}\ }\textbf {\bibinfo {volume} {167}},\
  \bibinfo {pages} {331} (\bibinfo {year} {1968})}\BibitemShut {NoStop}%
\bibitem [{\citenamefont {Sanna}\ and\ \citenamefont
  {Gross}(2014)}]{Sanna_Migdal}%
  \BibitemOpen
  \bibfield  {author} {\bibinfo {author} {\bibfnamefont {A.}~\bibnamefont
  {Sanna}}\ and\ \bibinfo {author} {\bibfnamefont {E.~K.~U.}\ \bibnamefont
  {Gross}},\ }\href@noop {} {} (\bibinfo {year} {2014}),\ \bibinfo {note} {to
  be published}\BibitemShut {NoStop}%
\bibitem [{Note4()}]{Note4}%
  \BibitemOpen
  \bibinfo {note} {We are actually not reporting the Eliashberg result but that
  coming from the Allen-Dynes (AD) formula. The reason for this choice is that
  the two approaches agree perfectly (the difference being less than 1\protect
  \tmspace +\thinmuskip {.1667em}K) for the phonon case. But in addition when
  including $\mu ^*$\ the AD formula depends only on it, while the Eliashberg
  equations also depend on the Coulomb frequency cut-off (that changes the
  meaning of the $*$ in $\mu ^*$. If we want to use a conventional value of
  $\mu ^*$\ between 0.1 and 0.15~\cite {AllenMitrovic1983} it is then better to
  use the parametrized AD version of the Eliashberg method}\BibitemShut
  {NoStop}%
\bibitem [{\citenamefont {Morel}\ and\ \citenamefont
  {Anderson}(1962)}]{MorelAnderson_1962}%
  \BibitemOpen
  \bibfield  {author} {\bibinfo {author} {\bibfnamefont {P.}~\bibnamefont
  {Morel}}\ and\ \bibinfo {author} {\bibfnamefont {P.~W.}\ \bibnamefont
  {Anderson}},\ }\href {\doibase 10.1103/PhysRev.125.1263} {\bibfield
  {journal} {\bibinfo  {journal} {Phys. Rev.}\ }\textbf {\bibinfo {volume}
  {125}},\ \bibinfo {pages} {1263} (\bibinfo {year} {1962})}\BibitemShut
  {NoStop}%
\bibitem [{\citenamefont {Suhl}\ \emph {et~al.}(1959)\citenamefont {Suhl},
  \citenamefont {Matthias},\ and\ \citenamefont
  {Walker}}]{SMW_multibandBCS_PRL1959}%
  \BibitemOpen
  \bibfield  {author} {\bibinfo {author} {\bibfnamefont {H.}~\bibnamefont
  {Suhl}}, \bibinfo {author} {\bibfnamefont {B.}~\bibnamefont {Matthias}}, \
  and\ \bibinfo {author} {\bibfnamefont {L.}~\bibnamefont {Walker}},\ }\href
  {http://journals.aps.org/prl/abstract/10.1103/PhysRevLett.3.552} {\bibfield
  {journal} {\bibinfo  {journal} {Physical Review Letters}\ }\textbf {\bibinfo
  {volume} {3}},\ \bibinfo {pages} {552} (\bibinfo {year} {1959})}\BibitemShut
  {NoStop}%
\bibitem [{\citenamefont {McMillan}(2002)}]{PaulMc_NatHPMat}%
  \BibitemOpen
  \bibfield  {author} {\bibinfo {author} {\bibfnamefont {P.~F.}\ \bibnamefont
  {McMillan}},\ }\href {\doibase http://dx.doi.org/10.1038/nmat716} {\bibfield
  {journal} {\bibinfo  {journal} {Nat. Mat.}\ }\textbf {\bibinfo {volume}
  {1}},\ \bibinfo {pages} {1476} (\bibinfo {year} {2002})}\BibitemShut
  {NoStop}%
\bibitem [{\citenamefont {Essenberger}\ \emph {et~al.}(2014)\citenamefont
  {Essenberger}, \citenamefont {Sanna}, \citenamefont {Linscheid},
  \citenamefont {Tandetzky}, \citenamefont {Profeta}, \citenamefont {Cudazzo},\
  and\ \citenamefont {Gross}}]{Frank_SF_PRB2014}%
  \BibitemOpen
  \bibfield  {author} {\bibinfo {author} {\bibfnamefont {F.}~\bibnamefont
  {Essenberger}}, \bibinfo {author} {\bibfnamefont {A.}~\bibnamefont {Sanna}},
  \bibinfo {author} {\bibfnamefont {A.}~\bibnamefont {Linscheid}}, \bibinfo
  {author} {\bibfnamefont {F.}~\bibnamefont {Tandetzky}}, \bibinfo {author}
  {\bibfnamefont {G.}~\bibnamefont {Profeta}}, \bibinfo {author} {\bibfnamefont
  {P.}~\bibnamefont {Cudazzo}}, \ and\ \bibinfo {author} {\bibfnamefont
  {E.~K.~U.}\ \bibnamefont {Gross}},\ }\href {\doibase
  10.1103/PhysRevB.90.214504} {\bibfield  {journal} {\bibinfo  {journal} {Phys.
  Rev. B}\ }\textbf {\bibinfo {volume} {90}},\ \bibinfo {pages} {214504}
  (\bibinfo {year} {2014})}\BibitemShut {NoStop}%
\bibitem [{\citenamefont {Akahama}\ \emph
  {et~al.}(1993{\natexlab{a}})\citenamefont {Akahama}, \citenamefont
  {Kobayashi},\ and\ \citenamefont {Kawamura}}]{Akahama_S_PRB1993}%
  \BibitemOpen
  \bibfield  {author} {\bibinfo {author} {\bibfnamefont {Y.}~\bibnamefont
  {Akahama}}, \bibinfo {author} {\bibfnamefont {M.}~\bibnamefont {Kobayashi}},
  \ and\ \bibinfo {author} {\bibfnamefont {H.}~\bibnamefont {Kawamura}},\
  }\href {\doibase 10.1103/PhysRevB.47.20} {\bibfield  {journal} {\bibinfo
  {journal} {Phys. Rev. B}\ }\textbf {\bibinfo {volume} {47}},\ \bibinfo
  {pages} {20} (\bibinfo {year} {1993}{\natexlab{a}})}\BibitemShut {NoStop}%
\bibitem [{\citenamefont {Akahama}\ \emph
  {et~al.}(1993{\natexlab{b}})\citenamefont {Akahama}, \citenamefont
  {Kobayashi},\ and\ \citenamefont {Kawamura}}]{Akahama_Se_PRB1993}%
  \BibitemOpen
  \bibfield  {author} {\bibinfo {author} {\bibfnamefont {Y.}~\bibnamefont
  {Akahama}}, \bibinfo {author} {\bibfnamefont {M.}~\bibnamefont {Kobayashi}},
  \ and\ \bibinfo {author} {\bibfnamefont {H.}~\bibnamefont {Kawamura}},\
  }\href {\doibase 10.1103/PhysRevB.48.6862} {\bibfield  {journal} {\bibinfo
  {journal} {Phys. Rev. B}\ }\textbf {\bibinfo {volume} {48}},\ \bibinfo
  {pages} {6862} (\bibinfo {year} {1993}{\natexlab{b}})}\BibitemShut {NoStop}%
\bibitem [{\citenamefont {Akahama}\ \emph {et~al.}(1997)\citenamefont
  {Akahama}, \citenamefont {Kobayashi},\ and\ \citenamefont
  {Kawamura}}]{Akahama_Se_met_PRB1997}%
  \BibitemOpen
  \bibfield  {author} {\bibinfo {author} {\bibfnamefont {Y.}~\bibnamefont
  {Akahama}}, \bibinfo {author} {\bibfnamefont {M.}~\bibnamefont {Kobayashi}},
  \ and\ \bibinfo {author} {\bibfnamefont {H.}~\bibnamefont {Kawamura}},\
  }\href {\doibase 10.1103/PhysRevB.56.5027} {\bibfield  {journal} {\bibinfo
  {journal} {Phys. Rev. B}\ }\textbf {\bibinfo {volume} {56}},\ \bibinfo
  {pages} {5027} (\bibinfo {year} {1997})}\BibitemShut {NoStop}%
\bibitem [{\citenamefont {Margine}\ and\ \citenamefont
  {Giustino}(2013)}]{Margine_anisoEliashberg_PRB2013}%
  \BibitemOpen
  \bibfield  {author} {\bibinfo {author} {\bibfnamefont {E.~R.}\ \bibnamefont
  {Margine}}\ and\ \bibinfo {author} {\bibfnamefont {F.}~\bibnamefont
  {Giustino}},\ }\href {\doibase 10.1103/PhysRevB.87.024505} {\bibfield
  {journal} {\bibinfo  {journal} {Phys. Rev. B}\ }\textbf {\bibinfo {volume}
  {87}},\ \bibinfo {pages} {024505} (\bibinfo {year} {2013})}\BibitemShut
  {NoStop}%
\end{thebibliography}%
\end{document}